\documentclass[aps, prb, twocolumn,amsmath,amssymb,reprint]{revtex4-2}
\usepackage{graphicx}
\usepackage{float} 
\usepackage{subfigure} 
\usepackage{dcolumn}
\usepackage{bm}
\usepackage[dvipsnames]{xcolor}
\usepackage{physics}
\usepackage{pifont}
\usepackage{comment}
\newcommand{\MY}{\textcolor{red}}

\usepackage{hyperref}
\begin{document}
\title{Basic formulation and first-principles implementation of nonlinear magneto-optical effects}
 
\author{Haowei \surname{Chen}$^{1}$}
\author{Meng \surname{Ye}$^{1}$}
\email{mengye@mail.tsinghua.edu.cn}
\author{Nianlong \surname{Zou}$^{1}$}
\author{Bing-lin \surname{Gu}$^{3}$}
\author{Yong \surname{Xu}$^{1,2,4,5,6}$}
\email{yongxu@mail.tsinghua.edu.cn}
\author{Wenhui \surname{Duan}$^{1,2,3,4,5}$}

\affiliation{$^{1}$State Key Laboratory of Low Dimensional Quantum Physics and Department of Physics, Tsinghua University, Beijing, 100084, China\\
$^2$Tencent Quantum Laboratory, Tencent, Shenzhen, Guangdong 518057, China\\
$^{3}$Institute for Advanced Study, Tsinghua University, Beijing 100084, China \\
$^{4}$Frontier Science Center for Quantum Information, Beijing, China\\
$^{5}$Beijing Academy of Quantum Information Sciences, Beijing 100193, China\\
$^{6}$RIKEN Center for Emergent Matter Science (CEMS), Wako, Saitama 351-0198, Japan}

\date{\today}
\begin{abstract}
First-principles calculation of nonlinear magneto-optical effects has become an indispensable tool to reveal the geometric and topological nature of electronic states and to understand light-matter interactions. While intriguingly rich physics could emerge in magnetic materials, further methodological developments are required to deal with time-reversal symmetry breaking, due to the degeneracy and gauge problems caused by symmetry and the low-frequency divergence problem in the existing calculation formalism. Here we present a gauge-covariant and low-frequency convergent formalism for the first-principles computation. Remarkably, this formalism generally works for both non-magnetic and magnetic materials with or without band degeneracy. Reliability and capability of our method are demonstrated by studying example materials (i.e.,  bilayers of MnBi$_2$Te$_4$ and CrI$_3$) and comparing with published results. Moreover, an importance correction term that ensures gauge covariance of degenerate states is derived, whose influence on physical responses is systematically checked. Our method enables computation of nonlinear magneto-optical effects in magnetic materials and paves the way for exploring rich physics created by the interplay of light and magnetism.
\end{abstract}
\maketitle

\section{Introduction}
Optical materials and phenomena play indispensable roles in modern science and technology.
Along with the development of laser technology, nonlinear optical materials in which the polarization responses nonlinearly to the electric field of light are discovered and nonlinear optical phenomena such as the bulk photovoltaic effect (BPVE) \cite{review_BPVE,LNO_BPVE,BTO_BPVE} and the second-harmonic generation (SHG) \cite{SHG_1961} have been widely investigated. 
The BPVE is the generation of a rectification current driven by light in a single-phase material lacking inversion symmetry and it originates from the shift current mechanism \cite{Rappe_BPVE}.
SHG, which is the generation of photons with twice the frequency of incident photons in non-centrosymmetric materials, is commonly used as a tool for symmetry characterization and laser frequency conversion. Remarkably, the intrinsic connections between optical responses and the quantum geometry in momentum space have been revealed recently, giving rise to anomalous optical responses in topological materials \cite{Nagaosa_NLO_Topology_2016,Nagaosa_NLO_geometry_2017,Nagaosa_NLO_geometry_topology_2020,Weyl_SC_2019, chiral_Weyl_theory,chiral_Weyl_expt}.

Light-matter interaction in magnetic materials brings up profound physics due to the interplay between magnetism and the electromagnetic wave.
In the linear regime, the plane of polarization can be rotated when light is reflected by
or transmitted through magnetic materials, which is called the linear magneto-optical Kerr effect or Faraday effect. 
In the nonlinear regime, the magnetization-sensitive rectification current has recently been proposed in antiferromagnetic (AFM) bilayers of CrI$_3$ and MnBi$_2$Te$_4$ with parity-time ($\mathcal{PT}$) symmetry \cite{Yan_CrI3,Qian_MBT,Yang_MBT}, whose non-magnetic counterparts are absent. 
Magnetization-sensitive SHG has long been a tool to probe surface and interface magnetization in metallic materials since 1990s \cite{Reif_MSHG_1991,Reif_MSHG_1993,Kirilyuk_MSHG_review_2002,Kirilyuk_MSHG_review_2005} and later used to investigate complex magnetic orders in bulk magnetic oxides, such as in Cr$_2$O$_3$ \cite{Fiebig_MSHG_Cr2O3}.
Due to the spectral and spatial resolution, SHG is an indispensable tool to detect magnetic symmetries, orders, domains and phase transitions, especially in low dimensions \cite{Fiebig_MSHG_review_2005}. 
A recent SHG measurement discovered a gigantic signal from AFM bilayer CrI$_3$ where the microscopic origin of the large response remains unclear \cite{Wu_CrI3_SHG}.

Theoretical calculations based on first-principles methods are powerful tools to reveal microscopic correlations between the energy spectrum, the momentum space geometry and optical effects. 
Moreover, nonlinear magneto-optical effects are especially powerful in studying AFM insulators with $\mathcal{PT}$ symmetry, where linear magneto-optical effects, such as Kerr and Faraday effects, are absent. 
However, several problems hinder the theoretical investigation of nonlinear magneto-optical effects.
Firstly, the conventional formalism of nonlinear magneto-optical effects is ill-defined in the presence of $\mathcal{PT}$ symmetry due to the Kramers-like degeneracy at each $k$ point in the band structure. 
The conventional expressions of nonlinear conductivity and electric susceptibility \cite{Qian_MBT,Yan_CrI3,Sipe_1993} only consider non-degenerate bands with an arbitrary phase freedom for eigenstates at each $k$ point, which is also called the U(1)-gauge freedom. 
However, due to the Kramers-like degeneracy, formulations under $\mathcal{PT}$ symmetry should be
invariant under an arbitrary 2 $\times$ 2 unitary transformation in the subspace spanned by the doubly degenerate eigenstates, corresponding to a U(2)-gauge freedom.
Although Watanabe and Yanase \cite{Watanabe} have proposed to directly replace the U(1)-covariant derivative by the U(2)-covariant derivative in the presence of $\mathcal{PT}$ symmetry for magnetization-sensitive shift current conductivity, detailed derivations are missing.

Secondly, the conventional expression of second-order electric susceptibility including SHG has a low-frequency divergent problem by including leading term factors of $\omega^{-1}$ and $\omega^{-2}$ for gapped insulators with the incident light frequency $\omega$ \cite{Sipe_1993,Sipe_1995,Sipe_2000}.
While the low-frequency divergence is frequently observed in physical properties of metallic materials due to the response from the Fermi surface \cite{Nagaosa_LowFreq_2020}, the low-frequency divergence is unphysical in gapped insulators.
Even if the expression has a $0/0$ form and is not divergent at $\omega \to 0$, the $0/0$ form expression   can cause sizable numerical instability in computation \cite{Sipe_1993,Sipe_1995,Sipe_2000}.
Fortunately, for SHG susceptibility, it has been shown that the divergence can be eliminated in the presence of time-reversal ($\mathcal{T}$) symmetry  \cite{Sipe_1991,Sipe_1993,Sipe_1995,Sharma_SHG_2003,Rashkeev_SHG_1998}. However, the divergence could be problematic for magnetic materials. 
The $\mathcal{T}$-symmetry simplified ($\mathcal{T}$-smp) expression in which $\mathcal{T}$ symmetry is only applied to the divergent terms have been implemented in many computational codes \cite{ABINIT,Chen_SHG_1999,Zhang_SHG_2014,Qian_SHG_2017,Yang_SHG_Perovskite,Guo_SHG_2005,Wang_NLO_Wannier} and reasonable agreement between theoretical predictions and experiments has been reached for non-magnetic materials \cite{Zhang_SHG_2020}. 
While $\mathcal{T}$-smp expressions can still be used to estimate the SHG susceptibility in magnetic materials \cite{Yang_CrI3_SHG}, the divergent terms that are unique to magnetic materials are missing. 
More importantly, the same divergence appears in the general second-order susceptibility (including sum-frequency generation and difference-frequency generation), which cannot be removed even under $\mathcal{T}$ symmetry. 
Therefore, it is imperative to solve the above mentioned problems to facilitate and advance the first-principles studies of nonlinear magneto-optical effects.

In this work, we presented general solutions to complete the computational methods for second-order magneto-optical effects in which the physical consequence of $\mathcal{PT}$ symmetry was derived and investigated, and a full-frequency convergent formalism of second-order susceptibility was proposed. 
The remainder of this article is organized as follows.
In Sec. \ref{sec:formula}, we derived the expressions for second-order magneto-optical responses based on density matrix perturbation method for degenerate bands and a full-frequency convergent formalism for second-order susceptibility including SHG. 
Sec. \ref{sec:symmetry} discussed the symmetry properties of second-order magneto-optical responses under $\mathcal{T}$ and $\mathcal{PT}$ operations, including charge/spin current and susceptibility. 
Sec. \ref{sec:methods} introduced the implementation of our methods in first-principles calculations and computational details. 
In Sec. \ref{sec:results}, we investigated magnetization-sensitive conductivity and susceptibility in two prototypical two-dimensional AFM materials with $\mathcal{PT}$ symmetry, which are bilayer CrI$_3$ and MnBi$_2$Te$_4$.

\section{Basic formulation} \label{sec:formula}

\subsection{General formulae with Kramers-like degeneracy}
The conventional formulae for second-order photo-current and electric polarization only consider non-degenerate bands, and therefore are only U(1)-gauge invariant \cite{Sipe_2000,Qian_MBT}.
However, in magnetic materials with $\mathcal{PT}$ symmetry, Kramers-like degeneracy appears at every $k$ point and the choice of eigenstates in the doubly degenerate subspace has a gauge freedom determined by a $2\times 2$ unitary matrix. 
As physical results are invariant in spite of an arbitrary unitary rotation in the degenerate subspace, the formulae for physical observables under $\mathcal{PT}$ symmetry should be U(2)-gauge invariant.
Clearly, certain terms are missing in the conventional formula in the presence of $\mathcal{PT}$ symmetry
and in what follows, we will demonstrate that the U(2)-invariant formulae can be retrieved with special treatment at degeneracy \cite{Sipe_2000}.

In the length gauge and under long-wavelength approximation, the electromagnetic wave is introduced through an $\mathbf{E}(t) \cdot \mathbf{\hat{x}}$ term to the unperturbed Hamiltonian $H_0$ where a monochromatic light at frequency $\omega$ is written as $\mathbf{E}(t) = \mathbf{E}(\omega)e^{-i\omega t} + $ c.c. 
The Hamiltonian in second quantization form can be found in Appendix \ref{appendix Basic Hamiltonian}.
The evaluation of the position operator $\mathbf{\hat{x}}$ is the key to the calculation of polarization and current responses, and the matrix elements of position operator can be written as \cite{Sipe_2000}
\begin{equation} 
\begin{split}
    &\langle n_{\mu} \mathbf{k}|\mathbf{\hat{x}}|m_{\nu}\mathbf{k^{\prime}}\rangle \\
    =& \delta_{nm}\delta_{{\mu}{\nu}}[\delta(\mathbf{k}-\mathbf{k^{\prime}})\mathbf{\xi}_{n_{\mu} m_{\nu}}(\mathbf{k}) + i\frac{\partial}{\partial\mathbf{k}}\delta(\mathbf{k}-\mathbf{k^{\prime}})]\\
    &+ (1-\delta_{nm}\delta_{{\mu}{\nu}})\mathbf{\xi}_{n_{\mu} m_{\nu}}(\mathbf{k})\delta(\mathbf{k}-\mathbf{k^{\prime}})\\
    =& ({\mathbf{\hat{x}}}_{\rm intra})_{n_{\mu} \mathbf{k},m_{\nu} \mathbf{k^{\prime}}}+({\mathbf{\hat{x}}}_{\rm inter})_{n_{\mu} \mathbf{k},m_{\nu} \mathbf{k^{\prime}}}\,.\\
\end{split} \label{eq:x_matrix}
\end{equation}
$|n_{\mu}\mathbf{k}\rangle = e^{i\mathbf{k}\cdot\hat{\mathbf{x}}}|u_{n_{\mu}}( \mathbf{k})\rangle $ is the eigenstate of $H_{\rm 0}$, 
and $|u_{n_{\mu}}( \mathbf{k})\rangle$ is the periodic part of the Bloch function.
We introduced two subscripts $n$ and $\mu$ for bands, where $n$ denotes bands with different energy, and $\mu$ is used in the subspace of degenerate bands. 
$\mu$ runs from 1 to 2 under $\mathcal{PT}$ symmetry, while the following derivations hold for an arbitrary $\mu$ value.  
The diagonal and off-diagonal parts of $\mathbf{\xi}_{n_{\mu} m_{\nu}}(\mathbf{k}) =i\langle u_{n_{\mu}} (\mathbf{k})|\nabla_{\mathbf{k}}|u_{m_{\nu}}( \mathbf{k})\rangle$ are 
the single band and two-band Berry connections $\mathbf{A}_{n_{\mu}}(\mathbf k)$ and $\mathbf{r}_{n_{\mu} m_{\nu}}(\mathbf k)$ defined as
\begin{equation} \label{eq: non-Abelian connections}
    \mathbf{\xi}_{n_{\mu} m_{\nu}}(\mathbf k) \equiv
    \begin{cases}
        \mathbf{A}_{n_{\mu}}(\mathbf k) & \text{if $n=m$ and ${\mu}={\nu}$}\,,\\
        \mathbf{r}_{n_{\mu} m_{\nu}}(\mathbf k) & \text{otherwise}\,.\\
    \end{cases}
\end{equation}
In the following formulae, we omit the $\mathbf{k}$ dependence for simplicity. 
The first (second) term of Eq.\eqref{eq:x_matrix} is the intraband (interband) position matrix element in an infinite crystal and the position operator matrix element between degenerate bands ($n=m, \mu\neq \nu$) are considered in the interband term.
As a result, the polarization can be divided into  $\mathbf{{P}}=e\mathbf{{x}}=\mathbf{{P}}_{\rm inter}+\mathbf{{P}}_{\rm intra}$ in which $e=-\abs{e}$ is the charge of an electron. 
Similarly, the current density along $a$ direction which is defined as the time derivative of electric polarization
is divided into two parts as 
\begin{equation} 
\begin{aligned}
    \langle \hat{j}^a(t) \rangle&=\frac{1}{i\hbar}\langle[\hat{P}^{a}_{\rm intra}(t),\hat{H}(t)]\rangle +\frac{d}{dt}\langle \hat{P}^a_{\rm inter}(t) \rangle\\
    &=\langle \hat{j}^a_{\rm intra}(t) \rangle + \frac{d}{dt} \langle \hat{P}^a_{\rm inter}(t) \rangle \,,
\end{aligned}\label{eq:jtot}
\end{equation}
where $\hat{H}(t)$ is the total Hamiltonian including the oscillating electric field. 

With the density matrix derived with perturbation expansion in Appendices \ref{appendix Basic Hamiltonian} and \ref{appendix: Basic density matrix}, the zeroth and first order responses including the finite frequency anomalous conductivity are derived in Appendix \ref{appendix: Basic responses}.
The second-order electric current and polarization with incoming electric field of frequency $\omega_{\beta}$ and $\omega_{\gamma}$ and the response frequency $\omega_{\Sigma} = \omega_{\beta} + \omega_{\gamma}$ are 
\begin{equation}\label{eq:Jintra}
\begin{aligned}
    &\langle \hat{j}_{\rm intra}^{a} (t)\rangle^{(2)}\\
    =&\frac{e^{3}}{{2}\hbar^{2} } \int [d\mathbf{k}] \left\{ E_{\beta}^{b} E_{\gamma}^{c} e^{-i \omega_{\Sigma} t} \left[ \frac{f_{nm} \Delta_{mn}^{a} r_{n_{\mu} m_{\nu}}^{c} r_{m_{\nu} n_{\mu}}^{b} }{\omega_{\Sigma}(\omega_{mn}-\omega_{\beta}-i/\tau)} \right.\right.\\ 
    &\left. \left. -\frac{f_{nm} r_{m_{\nu} n_{\mu}}^{b} r_{n_{\mu} m_{\nu};a}^{c}} {\omega_{mn}-\omega _{\beta}-i/\tau}\right]  + (b\beta \leftrightarrow c\gamma) \right\} \\
\end{aligned} 
\end{equation}
and 
\begin{equation} \label{eq:Pinter}
\begin{aligned}
    &\langle{\hat{P}}_{\rm inter}^{a} (t)\rangle^{(2)}\\
    =&{\frac{e^3}{2\hbar^2}}\int [d\mathbf{k}]  \sum_{n,l,m}\left[ \frac{[\ldots]E_{\beta}^{b} E_{\gamma}^{c} e^{-i\omega_\Sigma t}}{\omega_{ln}-\omega_{\Sigma}-i/\tau} + (b\beta \leftrightarrow c\gamma) \right]\,,
\end{aligned} 
\end{equation}
where the full expression of $\langle \hat{P}_{\rm inter}^{a}(t)\rangle^{(2)}$ can be found in Appendix \ref{appendix: Basic responses} and the Einstein summation convention is adopted in all formulae for repeated indices.
$\hbar$ is the reduced Planck's constant and $[d\mathbf{k}] = d\mathbf{k}/(2\pi)^d$ is the unit volume in reciprocal space of $d$ dimension.
An infinitesimal $1/\tau$ is introduced to avoid divergence at resonance, which can also be regarded as a relaxation rate. $a,b,c$ represent Cartesian directions, and $E_{\beta}$ is the shorthand notation for $E(\omega_\beta)$ which represents the amplitude of an electric field with frequency $\omega_\beta$. $(b\beta \leftrightarrow c\gamma)$ is the interchange of $b\beta$ and $c\gamma$ indices which reflects the fact that physical responses are irrelevant to the sequence of electric fields $E^b_\beta$ and $E^c_\gamma$ in the expression. $f_{nm}$, $\omega_{nm}$ and $\Delta_{mn}$ are the occupation, energy and group velocity differences between the $n$-th and the $m$-th bands. 
In gapped semiconductor at zero temperature, $f=1$ for valence band and $f=0$ for conduction band. 
$r_{n_{\mu} m_{\nu};a}^{b}=\partial_{k_a} r_{n_{\mu} m_{\nu}}^{b} - i(A^{a}_{n_{\mu}}r_{n_{\mu} m_{\nu}}^{b}-r^{b}_{n_{\mu} m_{\nu}}A^{a}_{m_{\nu}})$ is the U(1)-covariant derivative of $r_{n_{\mu} m_{\nu}}^{b}$, where the U(1)-covariant derivative of a non-degenerate eigenstate $|u_{n}(\mathbf{k})\rangle$ is $|u_{n}(\mathbf{k})\rangle_{;a}=\partial_{k_a} |u_{n}(\mathbf{k})\rangle+iA^a_{n}|u_{n}(\mathbf{k})\rangle=(1-|u_{n}(\mathbf{k})\rangle\langle u_{n}(\mathbf{k})|)\partial_{k_a} |u_{n}(\mathbf{k})\rangle\,$, which represents the projection of $\partial_{k_a} |u_{n\mathbf{k}}\rangle$ onto all states except itself.

It is convenient to use Fourier transformation to convert results into the frequency domain and define the conductivity and susceptibility in the frequency domain as 
\begin{equation}
\begin{aligned}
    j^a(\omega_\Sigma) &= \sigma^{abc}(-\omega_\Sigma; \omega_\beta, \omega_\gamma) E^b(\omega_\beta) E^c(\omega_\gamma) \,,\\
    P^a(\omega_\Sigma) &= \chi^{abc}(-\omega_\Sigma; \omega_\beta, \omega_\gamma) E^b(\omega_\beta) E^c(\omega_\gamma) \,.
\end{aligned}\label{eq:j-freq}
\end{equation}
In the following, we will write  $j_{\rm intra}^a(\omega)^{(2)}$ instead of $\langle \hat{j}^a_{\rm intra} (\omega)\rangle^{(2)}$ as abbreviation to represent the expectation value.

\subsection{Rectification current from degeneracy}
A changing electric polarization gives rise to a current, and the rectification current $(\omega_\beta = -\omega_\gamma = \pm \omega, \omega_\Sigma = 0)$ is of particular interest. 
If bands are non-degenerate which is equivalent to dropping the ${\mu},{\nu}$ indices and $\omega_{ln}\neq 0$ in Eq. (\ref{eq:Pinter}),  $\mathbf{P}_{\rm inter}(\omega_\Sigma=0)^{(2)}$ is finite, with $\omega_{\Sigma}\mathbf{P}_{\rm inter}(\omega_\Sigma=0)^{(2)}=0$, and $\mathbf{j}(\omega_\Sigma=0)^{(2)} =  \mathbf{j}_{\rm intra}(\omega_\Sigma=0)^{(2)} $ according to Eq. \eqref{eq:jtot}, which can reduce to the well-known U(1)-invariant formulae of rectification current densities \cite{Sipe_2000,Qian_MBT}.
However, if band degeneracy occurs, e.g. $\omega_{ln}=0$, degenerate bands contribute a polarization that is diverging as $\frac{1}{\omega_{\Sigma}}$ when $\omega_{\Sigma}$ approaches the static limit according to Eq. \eqref{eq:Pinter}.

The divergent polarization at $\omega_\Sigma \to 0$ corresponds to a DC current as 
\begin{equation} \label{eq:inter degenerate}
\begin{aligned}
    &\lim_{\omega_\Sigma \to 0} \left[ -i\omega_\Sigma   P^{a}_{\rm inter}(\omega_\Sigma)^{(2)} \right] \\
    =&\frac{ie^3}{{2}\hbar^2}\int [d\mathbf{k}] \left[ E_{\beta}^{b} E_{\gamma}^{c} \lim_{\omega_{\Sigma} \to 0}\left(\frac{-\omega_{\Sigma} }{-\omega_{\Sigma}-i/\tau}\right) f_{nm} r^b_{m_{\nu} n_{\mu}}  \right.\\
    &\left. \times  \frac{ r^{a}_{n_{\mu} n_{\lambda}}r_{n_{\lambda} m_{\nu}}^{c}-r^{c}_{n_{\mu} m_{\lambda}}r^{a}_{m_{\lambda} m_{\nu}}}{\omega_{m n}-\omega_{\beta}-i/\tau}   + {(b\beta \leftrightarrow c\gamma)} \right]\,,
\end{aligned}
\end{equation} 
which contains the Berry connection between degenerate bands, e.g. $r_{n_{\mu} n_{\lambda}}$.
The total rectification current is 
\begin{equation} 
\begin{aligned}
    & j^{a}(\omega_\Sigma=0)^{(2)} \\
    =& j^{a}_{\rm intra}(\omega_\Sigma=0)^{(2)}+ \lim_{\omega_\Sigma \to 0} \left[ -i\omega_\Sigma  P^{a}_{\rm inter}(\omega_\Sigma)^{(2)} \right] \\
    =&\frac{e^3}{{2}\hbar^2 } \int [d\mathbf{k}]\left\{ E_{\beta}^{b} E_{\gamma}^{c} \left[ \frac{f_{nm} \Delta_{m n}^{a} r_{n_{\mu} m_{\nu}}^{c} r_{m_{\nu} n_{\mu}}^{b}}{\omega_{\Sigma}(\omega_{m n}-\omega_{\beta}-i/\tau)} \right.\right.\\ 
    &-\left.\left. \frac{f_{n m}r_{m_{\nu} n_{\mu}}^{b} D^{a}[r_{n_{\mu} m_{\nu}}^{c}]}{\omega_{m n}-\omega _{\beta}-i/\tau}\right]  + {(b\beta \leftrightarrow c\gamma)} \right\}\\
\end{aligned}
\end{equation}
where ${D^a}[r_{n_{\mu} m_{\nu}}^{b}]=\partial_{k_a} r_{n_{\mu} m_{\nu}}^{b} - i\sum_{\lambda}(r^{a}_{n_{\mu} n_{\lambda}}r_{n_{\lambda} m_{\nu}}^{b}-r^{b}_{n_{\mu} m_{\lambda}}r^{a}_{m_{\lambda} m_{\nu}})$ is the U($N$)-covariant derivative of $r_{n_{\mu} m_{\nu}}^{b}$ and $N$ is the number of degeneracy determined by the index $\mu$. 
Similarly, the U($N$)-covariant derivative of degenerate eigenstates just projects $\partial_{k_a} |u_{n_{\mu}}(\mathbf{k})\rangle$ out of the $N$ dimensional degenerate subspace such that
$D^{a}|u_{n_{\mu}}(\mathbf{k})\rangle =\partial_{k_a} |u_{n_{\mu}}(\mathbf{k})\rangle+iA^a_{n_{\mu}}|u_{n_{\mu}}(\mathbf{k})\rangle 
+i\sum_{\nu \ne \mu}r^{a}_{n_{\nu}n_{\mu}}|u_{n_{\nu}}(\mathbf{k})\rangle
=(1-\sum_{\nu}|u_{n_{\nu}}(\mathbf{k})\rangle\langle u_{n_{\nu}}(\mathbf{k})|)\partial_{k_a} |u_{n_{\mu}}(\mathbf{k})\rangle \,$.
Up to now, we have presented a thorough derivation of the second-order electric current and have revealed the microscopic origin of the U($N$)-covariant derivative. 
In Appendix \ref{appendix: another way of U(2)}, we demonstrated an alternative derivation of the above formulae by including the position operator matrix between degenerate bands into the intraband term in Eq. \eqref{eq:x_matrix}.

In the intermediate region where the double degeneracy protected by $\mathcal{PT}$ symmetry is weakly broken with the energy splitting $E_{\rm deg}$, the fundamental formalisms in Eq. (\ref{eq:jtot} - \ref{eq:Pinter}) should be used to calculate the rectification current. 
As $E_{\rm deg}$ increases from 0 to finite, the rectification current turn to zero which is consistent with the U(1)-invariant result and an AC current peaked at $E_{\rm deg}$ appears.


Using the Sokhotski–Plemelj theorem, the denominator can be simplified as $(\omega_{mn}-\omega -i/\tau)^{-1} = {\rm P}(\omega_{mn}-\omega)^{-1} + i\pi \delta(\omega_{mn}-\omega)$ where P indicates principal part. Therefore, the total rectification current can be divided into four parts as normal injection current (NIC), magnetic injection current (MIC), normal shift current (NSC), and magnetic shift current (MSC)
\begin{equation}
\begin{aligned} 
    &j^{a}(\omega_\Sigma=0)^{(2)} \\
    =& \left[ \sigma^{abc}(0;\omega,-\omega) + \sigma^{abc}(0;-\omega,\omega)\right] \Re{E^{b}_\omega E^{c}_{-\omega}} \\
    &+i\left[\sigma^{abc}(0;\omega,-\omega) - \sigma^{abc}(0;-\omega,\omega) \right] \Im{E^{b}_\omega E^{c}_{-\omega}} \\
    =&2\left[\sigma^{abc}_{\rm NSC}(\omega)+\tau \eta^{abc}_{\rm MIC}(\omega) \right] \Re{E^{b}_\omega E^{c}_{-\omega}}\\
    &+2i\left[\sigma^{abc}_{\rm MSC}(\omega)+\tau\eta^{abc}_{\rm NIC}(\omega)\right] \Im{E^{b}_\omega E^{c}_{-\omega}} \,,\\
\end{aligned} \label{eq:JDC}
\end{equation}
with expressions
\begin{equation}\label{NSC}
\begin{aligned}
    \sigma_{\text{NSC}}^{a b c} (\omega)=&\frac{-i\pi e^{3}}{4\hbar^2} \int[d\mathbf{k}] f_{n m}\left(I_{mn}^{abc} + I_{mn}^{acb} \right) \\ 
    &\times [\delta\left(\omega_{n m}-\omega\right)+\delta\left(\omega_{m n}-\omega\right)] \,,
\end{aligned}
\end{equation}
\begin{equation}\label{MSC}
\begin{aligned} 
    \sigma_{\text{MSC}}^{a b c} (\omega)=&\frac{-i\pi e^{3}}{4\hbar^2} \int [d\mathbf{k}]  f_{n m}\left(I_{mn}^{abc} - I_{mn}^{acb} \right) \\ 
    & \times [\delta\left(\omega_{m n}-\omega\right)-\delta\left(\omega_{n m}-\omega\right)]\,,
\end{aligned}
\end{equation}
\begin{equation} \label{NIC}
{\eta}^{abc}_{\text{NIC}}(\omega)=-\frac{\pi e^3}{2\hbar^2}\int[d\mathbf{k}]f_{mn}\Delta^{a}_{mn}(-i\Omega^{bc}_{mn})\delta(\omega_{nm}-\omega)\,,
\end{equation}
\begin{equation} \label{MIC}
{\eta}^{abc}_{\text{MIC}}(\omega)=-\frac{\pi e^3}{2\hbar^2}\int[d\mathbf{k}]f_{mn}\Delta^{a}_{mn}(2g^{bc}_{mn})\delta(\omega_{nm}-\omega)\,.
\end{equation}
The integrand $I_{mn}^{abc}=\sum_{{\mu}{\nu}}r^{b}_{m_{\nu} n_{\mu}}D^{a}[r^{c}_{n_{\mu} m_{\nu}}]$, the two-band Berry curvature $\Omega^{bc}_{mn}=-2\sum_{{\mu}{\nu}}\Im{r_{m_{\nu}n_{\mu}}^{b}r_{n_{\mu}m_{\nu}}^{c}}$, the two-band quantum metric $g^{bc}_{mn}=\sum_{\mu \nu}\Re{r_{m_{\nu}n_{\mu}}^{b}r_{n_{\mu}m_{\nu}}^{c}}$ \cite{Nagaosa_NLO_geometry_2017}\cite{Nagaosa_NLO_geometry_topology_2020}\cite{Provost1980}, and the group velocity difference $\Delta_{mn}$ are all gauge invariant and their symmetry properties are listed in Table \ref{Tab:basic quantities}.
The above results are consistent with literature \cite{Qian_MBT,Sipe_2000}, and additionally we generalized the formulae of NSC and MSC to a U($N$)-invariant form in the presence of $N$-fold degeneracy.
\begin{table} 
	\renewcommand{\arraystretch}{1.5}
	\centering
\begin{tabular}{ccccc}
\hline \hline 
\quad & $I_{mn}^{abc}(\mathbf k)$  & $\Omega_{m n}^{ab}(\mathbf k)$ & $g_{mn}^{ab}(\mathbf k)$&$\Delta_{mn}^{a}(\mathbf k)$\\
\hline 
$\mathcal{P}$ & $-I_{mn}^{abc}(\mathbf{-k})$  &$\Omega_{m n}^{ab}(\mathbf{-k})$&  $g_{mn}^{ab}(\mathbf{-k})$ &$-\Delta_{mn}^{a}(\mathbf {-k})$\\
\hline
$\mathcal{T}$ & $-I_{nm}^{abc}(\mathbf{-k})$  &$-\Omega_{m n}^{ab}(\mathbf{-k})$&  $g_{mn}^{ab}(\mathbf{-k})$ &$-\Delta_{mn}^{a}(\mathbf {-k})$\\
\hline 
$\mathcal{PT}$&  $I_{nm}^{abc}(\mathbf k)$ & $-\Omega_{m n}^{ab}(\mathbf k)$& $g_{m n}^{ab}(\mathbf k)$&$\Delta_{mn}^{a}(\mathbf {k})$\\
\hline \hline
\end{tabular}
\caption{Transformation rules of basic gauge-invariant quantities under $\mathcal{P}$, $\mathcal{T}$ and $\mathcal{PT}$, respectively.}
	\label{Tab:basic quantities}
\end{table}

 \subsection{Low-frequency divergent problem of susceptibility}
The general formula for second-order polarization can be derived through $\mathbf{P}(\omega)^{(2)}  =  \mathbf{j}_{\rm intra}(\omega)^{(2)}/({-i\omega})  +  \mathbf{P}_{\rm inter}(\omega)^{(2)}$ in which Berry connection terms containing degenerate bands are already included in $\mathbf{P}_{\rm inter}(\omega)^{(2)}$ in the conventional expression. 
Therefore, the expressions of polarization and susceptibility are compatible with degenerate bands and no additional modifications are required to fulfill the gauge invariant requirement.

However, two terms in the susceptibility with leading factors $\omega_{\Sigma}^{-1}$ and $\omega_{\Sigma}^{-2}$ originated from $ \mathbf{j}_{\rm intra}^{(2)}$ give rise to a divergent problem in the $\omega_\Sigma \to 0$ limit with the divergent terms 
\begin{equation}
    \begin{split}
        &\chi_{\rm dvg}^{abc}(-\omega_\Sigma; \omega_\beta, \omega_\gamma)\\
        =& \frac{-ie^3}{2\hbar^2\omega_\Sigma} \int [dk] f_{nm} \left[ \left(\frac{r_{mn}^b r_{nm;a}^c}{\omega_{mn}-\omega_\beta} + \frac{r_{mn}^c r_{nm;a}^b}{\omega_{mn}-\omega_\gamma}\right) \right. \\
        & \left. - \frac{\Delta_{mn}^a}{\omega_\Sigma} \left(\frac{ r_{nm}^c r_{mn}^b}{\omega_{mn}-\omega_\beta}  + \frac{ r_{nm}^b r_{mn}^c}{\omega_{mn}-\omega_\gamma} \right) \right]\\
    \end{split}\label{eq:chi_dvg}
\end{equation}
where the imaginary part $i/\tau$ in the denominator is not written out explicitly. 
The divergence is unphysical in an insulator with a finite energy gap. 
Even if $\chi_{\rm dvg}$ take a $0/0$ form and the divergence can be removed eventually, the expression in Eq. \eqref{eq:chi_dvg} is unfriendly to numerical calculations at low frequency. 
This problem has been noticed in SHG susceptibility ($\omega_\beta = \omega_\gamma = \omega, \omega_\Sigma=2\omega$) by Sipe and Ghahramani \cite{Sipe_1993} and they discovered that the divergence in SHG can be removed by applying $\mathcal{T}$ symmetry. 
However, the simplification does not work in magnetic materials and as a results, the current SHG expressions might be problematic in studying magnetic materials.
Furthermore, applying $\mathcal{T}$ symmetry fails to remove the divergence in the general susceptibility in Eq. \eqref{eq:chi_dvg}.
Therefore, the second-order susceptibility for arbitrary frequencies still suffers from an unphysical divergent problem even for non-magnetic materials.

In what follows, we present an alternate expression for the divergent term which is given by 
\begin{equation}
    \begin{split}
        &\chi_{\rm dvg}^{abc}(-\omega_\Sigma; \omega_\beta, \omega_\gamma)\\
        =&  \frac{-ie^3}{2\hbar^2} \int [dk] \frac{f_{nm}}{\omega_{mn}} \left[ \left( \frac{\rho_\beta r_{mn}^b r_{nm;a}^c}{\omega_{mn}-\omega_\beta} + \frac{\rho_\gamma r_{mn}^c r_{nm;a}^b }{\omega_{mn}-\omega_\gamma} \right) \right. \\
        &\left. - \frac{\Delta_{mn}^a}{\omega_{mn}}\left( \frac{\rho_\beta^2 r_{mn}^b r_{nm}^c}{\omega_{mn} - \omega_\beta} + \frac{\rho_\gamma^2 r_{mn}^c r_{nm}^b}{\omega_{mn}-\omega_\gamma} \right) \right]\,,
    \end{split}\label{eq:nodvg}
\end{equation}
where $\rho_\beta = \omega_\beta / \omega_\Sigma$ and $\rho_\gamma = \omega_\gamma / \omega_\Sigma$. 
As $\omega_\Sigma$ is completely removed from the denominator, the above expression does not have divergence and numerical instability in the zero-frequency limit and can be applied to both non-magnetic and magnetic materials. 
Eq. \eqref{eq:nodvg} is derived from Eq. \eqref{eq:chi_dvg} by the following techniques without assuming $\mathcal{T}$ symmetry: exchange of dumpy band indices $n$ and $m$, integration by parts, and the equality $-\omega^{-1} \left[ (\omega_{mn}-\omega)^{-1} + (\omega_{nm}-\omega)^{-1} \right] = \omega_{nm}^{-1} \left[ (\omega_{mn}-\omega)^{-1} - (\omega_{nm}-\omega)^{-1}\right]$. 
Although the U(1)-derivative appears in the first term of Eq. \eqref{eq:nodvg}, the full susceptibility can be rearranged in a way that terms with Berry connection between degenerate bands are grouped with terms with the U(1)-covariant derivative to form the U($N$)-covariant derivative.

As SHG is one of the most commonly used tool for symmetry detection and frequency conversion, we take SHG as an example to discuss properties of the `divergent' terms. 
The full expression of SHG is given in the Appendix \ref{appendix: SFG and SHG} and the `divergent' terms are 
\begin{equation}
    \begin{split}
        &\chi_{\rm dvg}^{abc}(-2\omega;\omega,\omega) \\
        =&-\frac{ie^3}{4\hbar^2}\int [d\mathbf{k}] f_{n m}\left[\frac{r_{mn}^c r_{nm;a}^b + r_{mn}^b r_{nm;a}^c}{\omega_{m n}(\omega_{m n}-\omega)} \right. \\
        & \left. -\frac{\Delta^{a}_{m n}(r^{b}_{n m}r^{c}_{m n}+r^{c}_{n m}r^{b}_{m n})}{2\omega^{2}_{m n}(\omega_{m n}-\omega)}\right] \,.
    \end{split} \label{eq:SHG_nodvg}
\end{equation}
The first term in Eq. \eqref{eq:SHG_nodvg} survives under $\mathcal{T}$ symmetry and is included in the conventional SHG expression \cite{Sipe_1993,Sipe_1995,Sharma_SHG_2003,Rashkeev_SHG_1998,ABINIT,Chen_SHG_1999,Zhang_SHG_2014,Qian_SHG_2017,Yang_SHG_Perovskite,Guo_SHG_2005,Wang_NLO_Wannier}. 
In addition, the first term is purely real in the static limit.
However, the second term in Eq.\eqref{eq:SHG_nodvg} is present only when $\mathcal{T}$ symmetry is broken and therefore it is a unique term for magnetic materials. 
We denote the second term by $\chi_{\rm new}(-2\omega;\omega,\omega)$ as it has not been calculated previously.
Unexpectedly, we discovered that although $\chi_{\rm new}$ has a low-frequency `divergent' form in Eq. \eqref{eq:chi_dvg}, $\chi_{\rm new}(\omega=0)$ does not contribute to static SHG at all, which 
can be concluded by permutation of the dummy band indices in the summation in Eq. \eqref{eq:SHG_nodvg}.
Therefore, the low-frequency `divergent' term $\chi_{\rm new}$ can be observed away from low-frequency regime but negligible in low-frequency regime.

\section{symmetry Analysis}\label{sec:symmetry}
\subsection{Charge and spin rectification currents}
The coefficients of second-order electro-optical responses are third-rank tensors, and their nonzero and independent elements are determined by their magnetic point group \cite{Bilbao}.  
As the electric field, polarization and current are all odd under $\mathcal{P}$, the second-order electro-optical responses are only present in $\mathcal{P}$-breaking materials, according to Eq. \eqref{eq:j-freq}. 
Additionaly, special attentions are paid to $\mathcal{T}$ and $\mathcal{PT}$ symmetry and the symmetry properties of quantum metric, two-band Berry curvature, and {\it etc.} are summarized in Table \ref{Tab:basic quantities}. 
As a result, both NSC and NIC are even under $\mathcal{T}$ while MSC and MIC are even under $\mathcal{PT}$ symmetry according to Eqs. (\ref{NSC} - \ref{MIC}).  
Furthermore, the current coupled to $\Re{E^{b}(\omega)E^{c}(-\omega)}$ can be induced both by a linearly-polarized light (LPL) and a circularly-polarized light (CPL), while the current coupled to $\Im{E^{b}(\omega)E^{c}(-\omega)}$ can only be induced by CPL with the direction of current determined by the helicity of light \cite{Qian_MBT}\cite{Watanabe}.

Up to now, we have focused on the charge photocurrent which is the collective motion of both spin-up and spin-down electrons.
As electrons also carry the spin degree of freedom, the spin current can also exists.
Although it is still challenging to properly define the spin current in materials with spin-orbit coupling (SOC) as spin is not a good quantum number in this circumstance \cite{Niu_spincurrent}, the conventional definition $j^{ab}=1/2(v^a s^b+s^b v^a)$ with $v$ and $s$ the velocity and spin operators respectively is still appropriate without SOC \cite{Rappe_spin_shift_current}\cite{Yang_spin_current}.

We adopted the conventional definition to analyse the symmetry of spin current.
As spin is $\mathcal{T}$ odd and $\mathcal{P}$ even, the symmetry of spin current and charge current are opposite under $\mathcal{T}$ and  $\mathcal{PT}$ symmetry.
For instance, while the charge current from the NSC and NIC mechanisms are $\mathcal{T}$ even, the spin current from those mechanisms are $\mathcal{T}$ odd.
As a result, spin current from NSC and NIC mechanisms only exists in magnetic materials. 
The symmetry analysis of both charge and spin currents are summarized in Table \ref{Tab:symm}. 
Additionally, in whatever situation, the spin current and charge current are both present.
For instance, the charge NSC and spin MIC are present simultaneously in $\mathcal{T}$-symmetric materials under LPL. 
\begin{table}
	\renewcommand{\arraystretch}{1.3}
	\centering
\begin{tabular}{ccccc} 
\hline \hline
\quad & $\sigma_{\rm NSC}$ & $\sigma_{\rm MSC}$ & ${\eta}_{\rm NIC}$ & ${\eta}_{\rm MIC}$ \\
\hline Light & LPL & CPL & CPL & LPL\\
\hline Current & charge/spin &  charge/spin & charge/spin & charge/spin\\
\hline $\mathcal{T}$ & $\checkmark$/$\times$ & $\times$/$\checkmark$ &$\checkmark$/$\times$ & $\times$/$\checkmark$\\
\hline $\mathcal{PT}$ & $\times$/$\checkmark$ & $\checkmark$/$\times$ & $\times$/$\checkmark$ & $\checkmark$/$\times$\\
 \hline \hline
\end{tabular}
\caption{Existence of charge and spin current under $\mathcal{T}$ and $\mathcal{PT}$ symmetry. $\checkmark$ means present while $\times$ means absent. }
	\label{Tab:symm}
\end{table}

Although the four mechanisms listed in Table \ref{Tab:symm} can generate both charge and spin currents, the microscopic origins are different. 
$\mathcal{P}$ breaking is the precondition of second-order charge and spin currents and it usually happens in the crystallographic structure.
However, if it is the spin order instead of the geometric structure that breaks $\mathcal{P}$, the presence of SOC is imperative for a non-zero charge current but not necessary for generating spin current.

\subsection{Second-order susceptibility}
Despite the intricate expression of second-order susceptibility, certain symmetry conditions are fulfilled. 
Firstly, second-order susceptibility has the intrinsic permutation symmetry which states that $\chi_2^{abc}(-\omega_\Sigma;\omega_\beta,\omega_\gamma)$ is unchanged by the simultaneous interchange of its last two frequency arguments and its last two Cartesian as $\chi_2^{abc}(-\omega_\Sigma;\omega_\beta,\omega_\gamma) = \chi_2^{acb}(-\omega_\Sigma;\omega_\gamma,\omega_\beta)$.
The intrinsic permutation symmetry is introduced in Eqs. \eqref{eq:Jintra} and \eqref{eq:Pinter} as a convention. 
Secondly, when all frequencies are detuned from resonance and the small imaginary part in the denominator can be ignored, the nonlinear susceptibility possesses full permutation symmetry which states that all of the frequency arguments of the nonlinear susceptibility can be freely interchanged, as long as the corresponding Cartesian indices are interchanged simultaneously: 
$\chi_2^{abc}(-\omega_\Sigma;\omega_\beta,\omega_\gamma) = \chi_2^{bca}(\omega_\beta;\omega_\gamma,-\omega_\Sigma)$.
The full permutation symmetry can be deduced from a consideration of the form of the electromagnetic field energy within a lossless medium \cite{Boyd_book}.
Thirdly, in the static limit,  the non-zero components of the susceptibility tensor are all equal $\chi_2^{abc}(0;0,0) = \chi_2^{bca}(0;0,0) = \chi_2^{cab}(0;0,0)$, which is an extension of the full permutation symmetry in the static limit, also called the Kleinman’s symmetry.
Moreover, the static susceptibility is a purely real quantity with the expression given in Appendix \ref{appendix: SFG and SHG}.

Furthermore, the second-order susceptibility which is odd in $\mathcal{P}$ can be decoupled into two contributions, a non-magnetic response $\chi_{\rm N}$ which is even in $\mathcal{T}$ and a magnetic response $\chi_{\rm M}$ which is odd in $\mathcal{T}$ with their expressions given in Appendix \ref{appendix: SFG and SHG}. 
$\chi_{\rm N}$ reflects the inversion symmetry breaking from the lattice and charge density while $\chi_{\rm M}$ is sensitive to the inversion symmetry breaking by the magnetic ordering.
The principal part and $\delta$-function part of the non-magnetic $\chi_{\rm N}$ are purely real and imaginary respectively, while the real and imaginary parts are opposite in the magnetic $\chi_{\rm M}$ \cite{Pershan_1963,Shen_1989_MSHG}.  
In other words, $\chi_{\rm N}$ is purely real and $\chi_{\rm M}$ is purely imaginary away from resonance. 
As the susceptibility in the static limit is purely real, $\chi_{\rm M}$ does not contribute to the static SHG.

Generally, the imaginary part of the susceptibility is responsible for energy dissipation of the electromagnetic field. 
However, in magnetic materials, $\chi_{\rm M}$ is imaginary even away from resonance, therefore, the meaning of dissipation needs to be scrutinized.
The derivation of energy dissipation from the first-order process is in Appendix \ref{appendix:dissipation}. 
The dissipation rate of the electromagnetic energy due to the second-order polarization $\mathbf{P}^{(2)}$ is given by
\begin{equation}
\begin{split}
    \mathbf{E}\cdot\frac{d}{dt} \mathbf{P}^{(2)}=&-2i[\omega_\Sigma\chi^{abc}_{2}(-\omega_\Sigma;\omega_\beta,\omega_\gamma)\\
    &- \omega_\beta\chi^{bac}_{2}(\omega_\beta;-\omega_\Sigma,\omega_\gamma) \\ &-\omega_\gamma\chi^{cab}_{2}(\omega_\gamma;-\omega_\Sigma,\omega_\beta)]E^{a}_{(-\Sigma)} E^{b}_\beta E^{c}_\gamma\,. 
\end{split}
\end{equation}
For frequencies detuned from resonance, the full-permutation symmetry of $\chi_2$ guarantees that the term in the bracket is zero. 
Therefore, only the $\delta$-function part of susceptibility participates in the dissipation no matter whether it is real or imaginary.

Moreover, $\chi_{\rm N}$ and $\chi_{\rm M}$ also corresponds to different measurable quantities. 
The sum-frequency generation (SFG) intensity ($I$) measured in experiments can be decoupled into a non-magnetic term, a magnetic term and interference terms as 
\begin{equation}
\begin{split}
    I =& |\chi_{\rm N}^{abc} E^b E^c + \chi_{\rm M}^{abc} E^b E^c|^2 \\
    =& |\chi_{\rm N}^{abc} E^b E^c|^2 + |\chi_{\rm M}^{abc} E^b E^c|^2 \\
    &+ (\chi_{\rm N}^{abc} E^b E^c) (\chi_M^{aij} E^i E^j)^* \\
    &+ (\chi_{\rm N}^{abc} E^b E^c)^* (\chi_M^{aij} E^i E^j)\,.
\end{split}
\end{equation}
For materials with $\mathcal{T}$ or $\mathcal{PT}$ symmetry, the SFG intensity is proportional to either the $\chi_{\rm N}$ term or the $\chi_{\rm M}$ term. 
When both $\mathcal{T}$ and $\mathcal{PT}$ symmetries are broken, e.g. in a polar ferromagnetic, the appearance of the third and the fourth terms reflects the interference between the magnetic and non-magnetic signals and the sign of the interference is determined by the the direction of the magnetic and the polar orders. 
As a result, domains with opposite magnetic/polar orders could show different SFG intensities which makes SFG an important tool for multiferroics detection.

\section{Computational methods} \label{sec:methods}
\subsection{Implementation of U(2)-covariant derivative} 
In the following, we present the implementation of the U(2)-invaraint formulae using both orthogonal and non-orthogonal basis. 
The general derivative is usually evaluated through a sum-over state method \cite{Sipe_2000,Moore_design_principle} or the Wannier interpolation method \cite{Wang_NLO_Wannier,Souza_Wannier_SC}. 
We adopted the Wannier interpolation method proposed in Ref. \cite{Wang_NLO_Wannier} for orthogonal basis and the scheme proposed in Ref. \cite{Wang_NLO_nonorthogonal} for non-orthogonal basis, while the Berry connection in both methods is evaluated based on non-degenerate perturbation theory.
Although methods with orthogonal and non-orthogonal basis sets have different gauge choices, as the optical responses are gauge invariant, the two methods are equivalent in principle. 
While the Wannier interpolation method is more computationally efficient, the method with non-orthogonal basis is more suitable for high-throughput calculations \cite{Wang_NLO_nonorthogonal}.
For SHG calculations, the three-band summation in Eq. (\ref{eq: full SHG}) requires a large number of bands to achieve convergence, therefore, the method with non-orthogonal basis is a more suitable choice to evaluate SHG responses.

We generalized the above mentioned methods by including degenerate perturbation to evaluate the Berry connection and the general derivative, which are unavoidable in the presence of $\mathcal{PT}$ symmetry.
For the Wannier interpolation method, we chose a specific gauge which satisfies that $(U^{\dagger}(\mathbf k_0)U(\mathbf k_0 + \delta \mathbf k))_{n_{\mu} n_{\nu}}$ is purely real for degenerate bands $n_{\mu}$ and $n_{\nu}$. 
Therefore, the non-Abelian connection $U^{\dagger}\partial_{k_a}U$ between two degenerate states vanishes, while the non-Abelian connections between non-degenerate states can be calculated as usual \cite{Wang_NLO_Wannier}
$$
(U^{\dagger} \partial_{k_a} U)_{n_{\mu} m_{\nu}} =
\begin{cases}
    -\frac{(U^{\dagger}\partial_{k_a}(H^{\rm (W)})U)_{n_{\mu} m_{\nu}}}{E_{n}-E_{m}} & E_n\ne E_m \\
    0 & E_n= E_m\\
\end{cases},
$$
where $H^{(W)}$ is the Hamiltonian matrix in the representation of local Wannier functions and $U$ is a matrix of which each column is an eigenvector of $H^{(W)}$. 
For the method of non-orthogonal basis, we again chose a specific gauge so that the matrix elements between degenerate bands in Ref. \cite{Wang_NLO_nonorthogonal} are modified as 
$$
v_{n_{\mu}}^{\dagger}S\partial_{k_a} v_{n_{\nu}}= -\frac{1}{2}v_{n_{\mu}}^{\dagger}(\partial_{k_a}  S)v_{n_{\nu}}\,.
$$
Here $S$ is the overlap matrix of the non-orthogonal basis, $v$ is a matrix with each column an generalized eigenvector of $H^{(W)}$ which is in the representation of a set of complete but not orthogonal basis. 
More details can be found in Ref. \cite{Wang_NLO_nonorthogonal}.

\subsection{First-principles methods}
First-principles calculations were performed both by Vienna $Ab \ initio$ Simulation Package (VASP) \cite{VASP} which is a plane-wave basis package and by OpenMX which is a pseudo-atomic basis package \cite{Openmx2003,Openmx2004}. The exchange-correlation functional was parameterized in the PBE form \cite{PBE}. PAW pseudopotentials \cite{PAW} and norm-conserving pseudopotentials \cite{NCpotential} were used in VASP and OpenMX, respectively.
We have included the effect of SOC.
For the 3$d$ orbitals in magnetic ions Mn and Cr, the Hubbard $U$ of 4\,eV and 3\,eV were used, respectively \cite{HubbardU}. 
For layered materials, we used DFT-D3 form van der Waals correction without damping \cite{DFT-D3}.
In VASP, the cut-off energy of plane waves was set to 350\,eV and 450\,eV for MnBi$_2$Te$_4$ and CrI$_3$, respectively. 
In OpenMX, Cr6.0-s3p3d2 and I9.0-s3p3d2f1 were chosen as our basis for Cr and I.
The convergence criterion for force and electronic calculations were 10\,meV/\AA\,and $10^{-6}$\,eV.
$k$-point samplings of $15\times15\times1$ and $13\times13\times1$ were used for MnBi$_2$Te$_4$ and CrI$_3$ in VASP, and a $7\times7\times1$ $k$-mesh was adopted for CrI$_3$ in OpenMX.

After getting the converged electronic structures, we either generated the maximally localized Wannier functions using Wannier90 \cite{wannier90} or used the non-orthogonal atomic basis from OpenMX to build the tight-binding (TB) Hamiltonian and calculate the optical responses \cite{Wang_NLO_Wannier,Wang_NLO_nonorthogonal}. 
We obtained 100 maximally localized orbitals for MnBi$_2$Te$_4$ and 112 maximally localized orbitals for CrI$_3$.   
In the calculation of MIC and MSC, we adopted a $400\times 400\times1$ kmesh.
For SHG susceptibility, the relaxation rate $\hbar/\tau$ was set to be 0.05\,eV, and the $k$-mesh was set to be $50\times50\times1$ which leads to identical results as $100\times100\times1$ $k$-point sampling.
The degenerate perturbation was applied when the energy difference between two bands are smaller than $E_{\rm deg}$ which was set to be 0.5\,meV in our calculations.

\begin{figure}
    \includegraphics[width=0.8\linewidth]{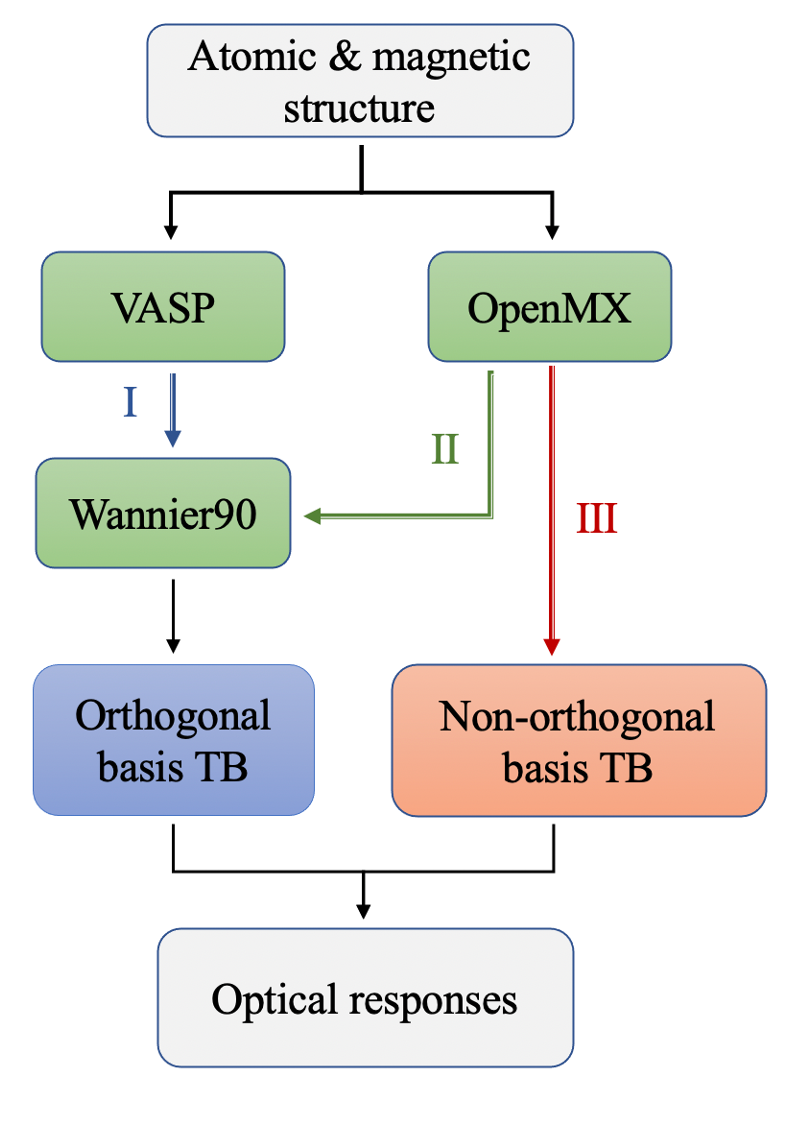}
    \caption{Calculation workflow of optical responses with three distinct routes. 
    In route \uppercase\expandafter{\romannumeral1}, after first-principles calculations by VASP, a TB Hamiltonian in orthogonal basis is constructed using Wannier90. For route \uppercase\expandafter{\romannumeral2}, the difference from route \uppercase\expandafter{\romannumeral1} is that DFT calculations are done by OpenMX.  In route \uppercase\expandafter{\romannumeral3}, a TB Hamiltonian in non-orthogonal basis is constructed directly from OpenMX.} 
 \label{Fig1}
\end{figure}

The flowchart of our computations including the electronic structure and optical response calculations is illustrated in Fig. \ref{Fig1}. 
Three different routes were adopted to calculate optical responses. 
In route \uppercase\expandafter{\romannumeral1}, the electronic structure is calculated by VASP and then maximally localized Wannier functions are constructed to obtain the TB Hamiltonian in an orthogonal basis set. 
The difference between route \uppercase\expandafter{\romannumeral2} and \uppercase\expandafter{\romannumeral1} is that the electronic structure is calculated by OpenMX rather than VASP. 
In route \uppercase\expandafter{\romannumeral3}, after the self-consistent calculation by OpenMX, we directly get the TB Hamiltonian in non-orthogonal basis.
The purpose of using multiple packages (VASP and OpenMX) and different basis sets (orthogonal and non-orthogonal) for TB Hamiltonian is to examine the influence of different electronic structures and gauge choices for optical responses separately. 

\section{Example studies and Results} \label{sec:results}
\subsection{Candidate materials}
\begin{figure}
    \includegraphics[width=\linewidth]{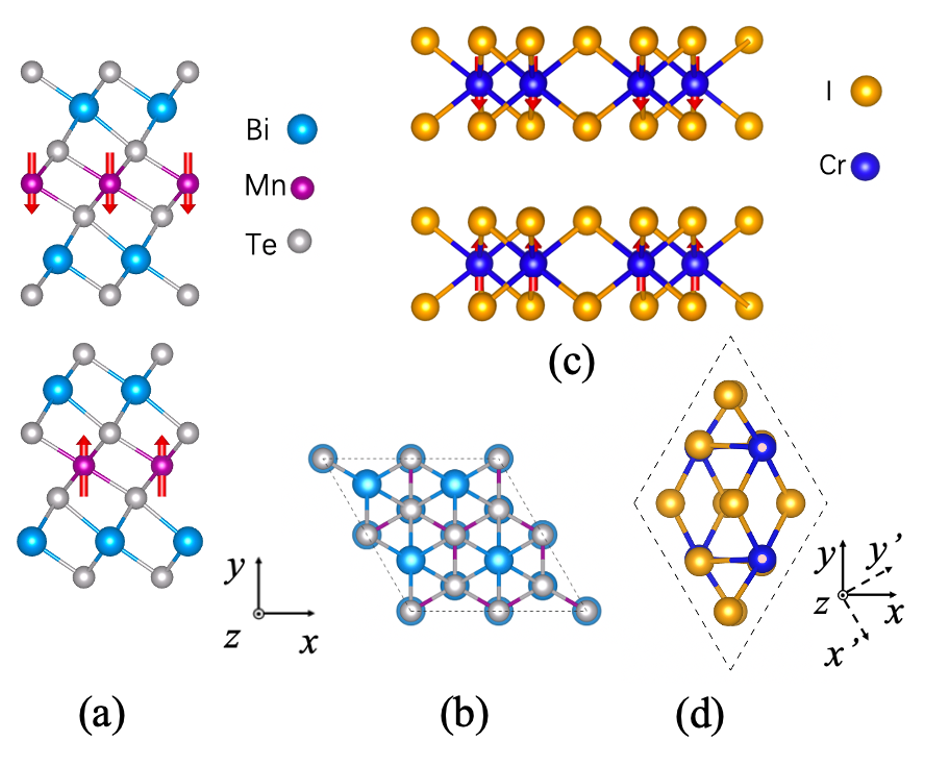} 
    \caption{Atomic and magnetic structures of bilayer MnBi$_2$Te$_4$ and CrI$_3$. (a, b) The side and top view of bilayer AFM-$z$ MnBi$_2$Te$_4$. (c, d) The side and top view of AFM-$z$ bilayer CrI$_3$. The Cartesian coordinates adopted in our calculations are marked by solid lines and the one adopted in Ref. \cite{Guo_CrI3} are in the dashed line.} \label{Fig2}
\end{figure}

We considered two prototypical materials: bilayer MnBi$_2$Te$_4$ \cite{Li_MBT} and bilayer CrI$_3$ \cite{Xiao_CrI3_DFT}, both of which have interlayer A-type AFM magnetic orders.
Both materials exhibit $\mathcal{PT}$ symmetry, which implies that the spatial inversion symmetry is broken only through the AFM spin order between two layers while the geometric structure is centrosymmetric as illustrated in Fig. \ref{Fig2}.
In this case, only magnetic responses including MIC, MSC and magnetization-sensitive SHG are present while normal responses NIC, NSC and non-magnetic SHG are forbidden, which makes them a simple platform to investigate the nonlinear magneto-optical responses.
Additionally, both materials are feasible in experiments \cite{Wu_CrI3_SHG,Xiao_CrI3_DFT} and their MIC has been studied theoretically in literature\cite{Yan_CrI3,Qian_MBT,Yan_CrI3,Guo_CrI3}, which provides valuable results for comparison.
For bilayer MnBi$_2$Te$_4$, we considered AB-type stacking (same stacking pattern as in bulk) in which each layer laterally shifted by [1/3, 1/3, 0].
For bilayer CrI$_3$, multiple stacking orders have been investigated previously \cite{Xiao_CrI3_DFT}, and we focused on the one with AB$^\prime$ stacking in which each layer laterally shifted by [1/3, 0, 0] as both MIC and MSC are allowed.
The stacking order, magnetic order, magnetic symmetry and independent tensor components are summarized in Table \ref{materials}.\\
\begin{table*} 
	\renewcommand{\arraystretch}{1.5}
	\centering
    \begin{tabular}{ccccccc}
        \hline \hline 
        Material & Stacking &  Magnetic point group & Magnetic order & MIC& MSC & MSHG\\ \hline 
        Bilayer MnBi$_2$Te$_4$ &  AB   &  $\bar{3}^{\prime}m^{\prime}$ & AFM-$z$ & $xxx$ &  none &$xxx$ \\ \hline 
        Bilayer CrI$_3$ &  AB$^{\prime}$ & $2/m^{\prime}$ & AFM-$z$ &$xxy$, $yxx$, $yyy$ &$xxy$ &$xxy$, $yxx$, $yyy$\\\hline \hline
    \end{tabular}
    \caption{The stacking order, magnetic point group, magnetic order, and independent tensor components of MIC, MSC and MSH in bilayer MnBi$_2$Te$_4$ and bilayer CrI$_3$.}
	\label{materials}
\end{table*}

\subsection{Magnetic injection current}
The MIC of bilayer MnBi$_2$Te$_4$ and bilayer CrI$_3$ have been calculated recently \cite{Qian_MBT,Guo_CrI3}. 
Although previous works haven't considered the U(2)-gauge problem under $\mathcal{PT}$ symmetry, fortunately, MIC is not subjected to modifications in $\mathcal{PT}$ symmetric case as demonstrated in Eq. \eqref{MIC}. 
In the following, we performed calculations and compared with previous works to validate our method and results.

\begin{figure}
    \includegraphics[width=\linewidth]{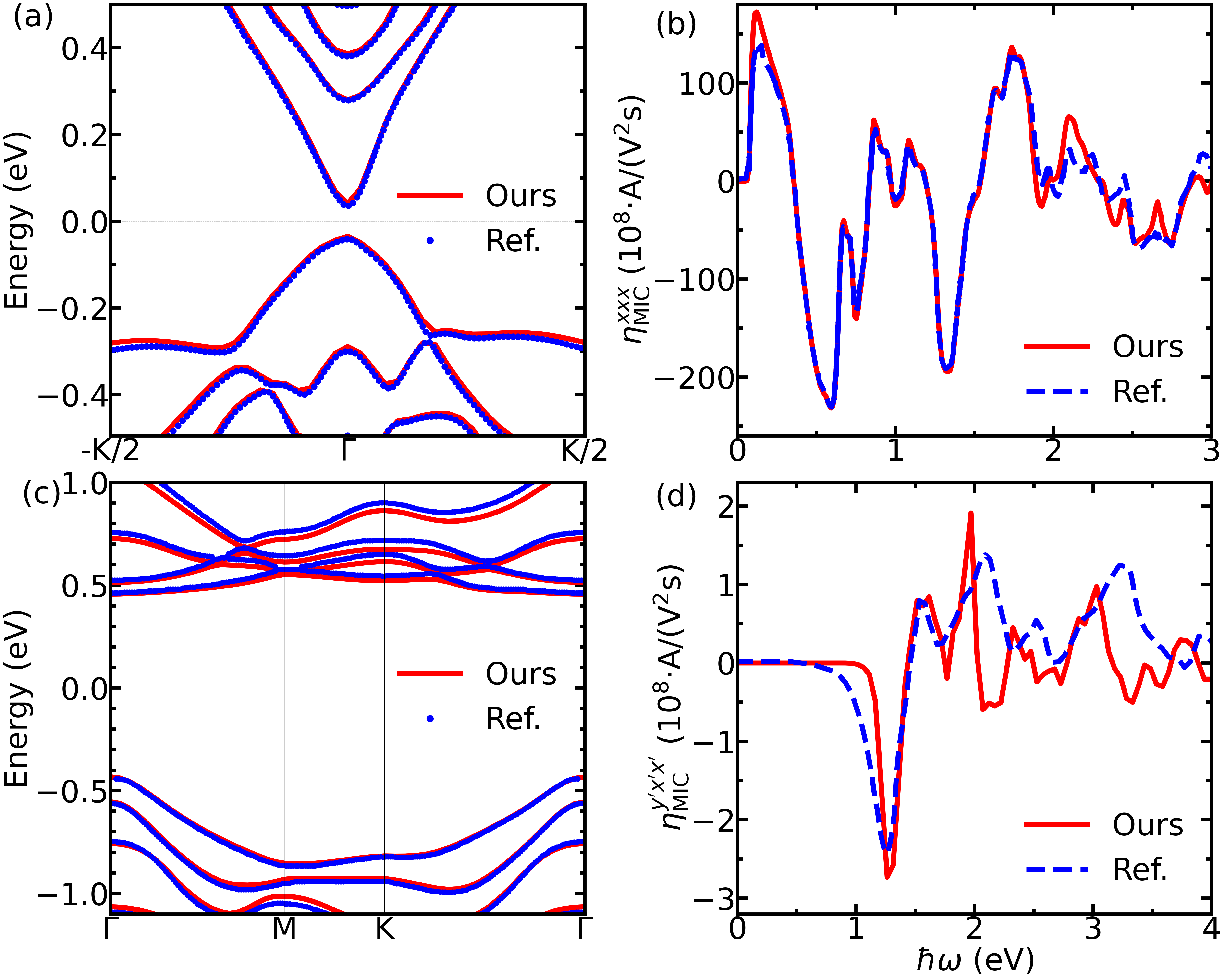} 
    \caption{Band structure and MIC conductivity of bilayer MnBi$_2$Te$_4$ and CrI$_3$. (a, b) Band structure and MIC conductivity of MnBi$_2$Te$_4$. Our results are in red solid line and  the results in Ref. \cite{Qian_MBT} are in blue dotted line. (c, d) Band structure and MIC conductivity of CrI$_3$. Our results are in red solid line, while the results in Ref. \cite{Guo_CrI3} are in blue dashed line. The coordinate was set to be the same as Ref. \cite{Guo_CrI3}, as illustrated in Fig. \ref{Fig2}. 
    }
    \label{Fig3}
\end{figure}

In bilayer MnBi$_2$Te$_4$, its interlayer coupling is AFM and the intralayer coupling is FM. 
The energy band is highly dispersive near $\Gamma$ point with a small gap of 0.076\,eV at $\Gamma$ point as shown in Fig. \ref{Fig3}(a) .
In addition, the comparison between our band structure (red solid line) and the one from Ref. \cite{Qian_MBT} (blue dotted line) along high symmetry line showed excellent agreement near band edge. 
Fig. \ref{Fig3}(b) shows the MIC conductivity of bilayer MnBi$_2$Te$_4$ along $xxx$ direction. 
Our result again shows an excellent agreement with the Ref. \cite{Qian_MBT}.

For bilayer CrI$_3$, in order to compare the band structure and MIC, we adopted the same Hubbard $U= 1$\,eV and the same Cartesian coordinates shown by dashed line in Fig. \ref{Fig2}(d) as Ref. \cite{Guo_CrI3}.
Fig. \ref{Fig3}(c, d) shows the calculated band structure and MIC conductivity. 
The band gap of bilayer CrI$_3$ is 0.89\,eV at $\Gamma$ point in our calculation and 0.78\,eV in the reference, therefore we applied a 0.11\,eV scissor operation to results in the reference to align the band gap in Fig. \ref{Fig3}(c) to 0.89\,eV. 
While the shape of bands is similar near band edge, noticeable differences exist in many bands.
In addition, the magnetic moment of Cr atom along $z$ axis is 3.12 $\mu_{B}$ in our calculation, while 3.21 $\mu_{B}$ in the reference. 
As we adopted the same parameters (Hubbard $U$, exchange-correlation functional, pseudopotentials, Van der Waals correction) in the electronic structure calculation, the above discrepancies might result from small differences in atomic structures.
For the MIC shown in Fig. \ref{Fig3}(d), the main characters of our results (red solid line) are similar to the reference (blue dashed line) including the location and height of the first peak.
However, due to discrepancies in band structures away from the band edge, the location and height of other peaks are shifted and modified.  
Therefore, combining the results of bilayer MnBi$_2$Te$_4$ and CrI$_3$, we can conclude that we are able to reproduce MIC results in literature while the detailed features of MIC are sensitive to geometric and electronic structures.

\subsection{Magnetic shift current}
The correction to ensure U(2)-gauge invariant is essential in the presence of $\mathcal{PT}$ symmetry which hasn't been considered in the previous computational works. 
Here we calculated the MSC conductivity of AB$^{\prime}$ stacking bilayer CrI$_3$ through three different routes shown in Fig. \ref{Fig1} and alternating between U(1)-covariant derivative and U(2)-covariant derivative to demonstrate the influence of the gauge choice.
As the results are very sensitive to details in electronic structure (see Fig. \ref{Fig7} in the Appendix), we stay with the electronic structure obtained from OpenMX for comparison.

\begin{figure*}
    \includegraphics[width=\linewidth]{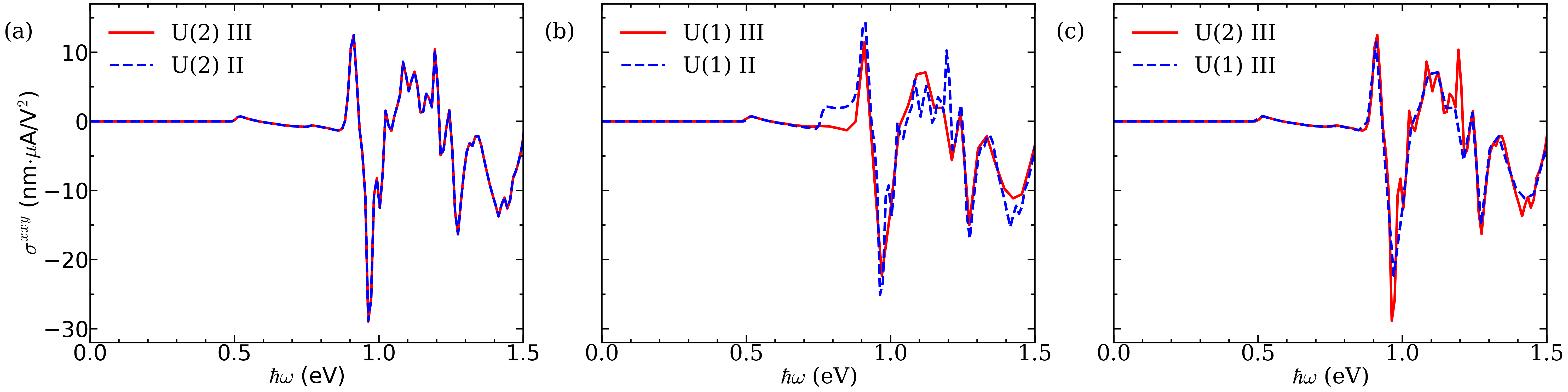}
    \caption{Magnetic shift current of bilayer CrI$_3$ calculated by (a) route \uppercase\expandafter{\romannumeral2} (blue dashed line) compared with route \uppercase\expandafter{\romannumeral3} (red solid line), both using the U(2)-invariant formula. The gap has been scissored to the same value. (b) The result of non-orthogonal basis (red solid line, route \uppercase\expandafter{\romannumeral3}) compared with that of orthogonal basis (blue dashed line, route \uppercase\expandafter{\romannumeral2}), both using U(1)-invariant formula.  (c) Results of U(2)-invariant formula (red solid line) versus U(1)-invariant formula (blue dashd line), both using non-orthogonal basis (route \uppercase\expandafter{\romannumeral3}).} \label{Fig4}
\end{figure*}

Figure \ref{Fig4}(a) shows the results from orthogonal basis method (blue dashed line) versus non-orthogonal basis method (red solid line) (route \uppercase\expandafter{\romannumeral2} vs route \uppercase\expandafter{\romannumeral3} in Fig. \ref{Fig1}) with the same U(2)-gauge-invariant formula. 
Since the two methods have different gauge choices, the excellently agreed results in Fig. \ref{Fig4}(a) confirm that our formula of MSC is truly gauge independent.

In contrast, Fig. \ref{Fig4}(b) shows the results of orthogonal basis method (blue dashed line) versus non-orthogonal basis method (red solid line) (route \uppercase\expandafter{\romannumeral2} versus route \uppercase\expandafter{\romannumeral3}) both using U(1)-form formula adopted in previous works. 
Although the same electronic structure guarantees that the peak positions resulted from the delta function in Eq. \eqref{MSC} are exactly the same, differences in peak height are prominent, especially at low-frequency region. 
For instance, they differ by more than three times at 1.2\,eV and there is a shoulder in the red solid line at 0.8\,eV while absent in the blue dashed line.
More importantly, the differences demonstrate that the previous formula with U(1)-covariant derivative is gauge dependent and the results are not reproducible due to different gauge choices.  
Therefore, the previous MSC formula with U(1)-covariant derivative cannot be used to describe physical observables in $\mathcal{PT}$-symmetric materials.

Furthermore, Fig. \ref{Fig4}(c) shows results from U(1)-invariant formula (blue dashed line) versus U(2)-invariant formula (red solid line) using non-orthogonal basis (both are in route \uppercase\expandafter{\romannumeral3}). 
Noticeable differences are observed at peak values due to the improper implementation of the covariant derivative in $\mathcal{PT}$-symmetric materials, implying the necessity of using U(2)-invariant formula.
%
%

\subsection{Magnetic second-harmonic generation}
In terms of the SHG response in magnetic materials, there are three sets of formulas that are available to use: the `DVG' expression which is the original expression in Ref. \cite{Sipe_1993} with low-frequency divergent problem shown in Eq. \eqref{eq:chi_dvg}, the `$\mathcal{T}$-SMP' expression in which the low-frequency divergent terms are simplified by $\mathcal{T}$ symmetry and have been implemented in many calculation packages \cite{ABINIT,Chen_SHG_1999,Zhang_SHG_2014,Qian_SHG_2017,Yang_SHG_Perovskite,Guo_SHG_2005,Wang_NLO_Wannier}, and the full-frequency convergent expression `CVG' proposed in this paper as demonstrated in Eq. \eqref{eq:nodvg}. 
As the gigantic SHG susceptibility in AB$^\prime$ CrI$_3$ has been measured \cite{Wu_CrI3_SHG} and calculated using the $\mathcal{T}$-SMP expression recently \cite{Yang_CrI3_SHG},
we compare the SHG response of bilayer CrI$_3$ along $xxy$ direction calculated using the DVG, $\mathcal{T}$-SMP and CVG expressions.

\begin{figure*}
    \includegraphics[width=\linewidth]{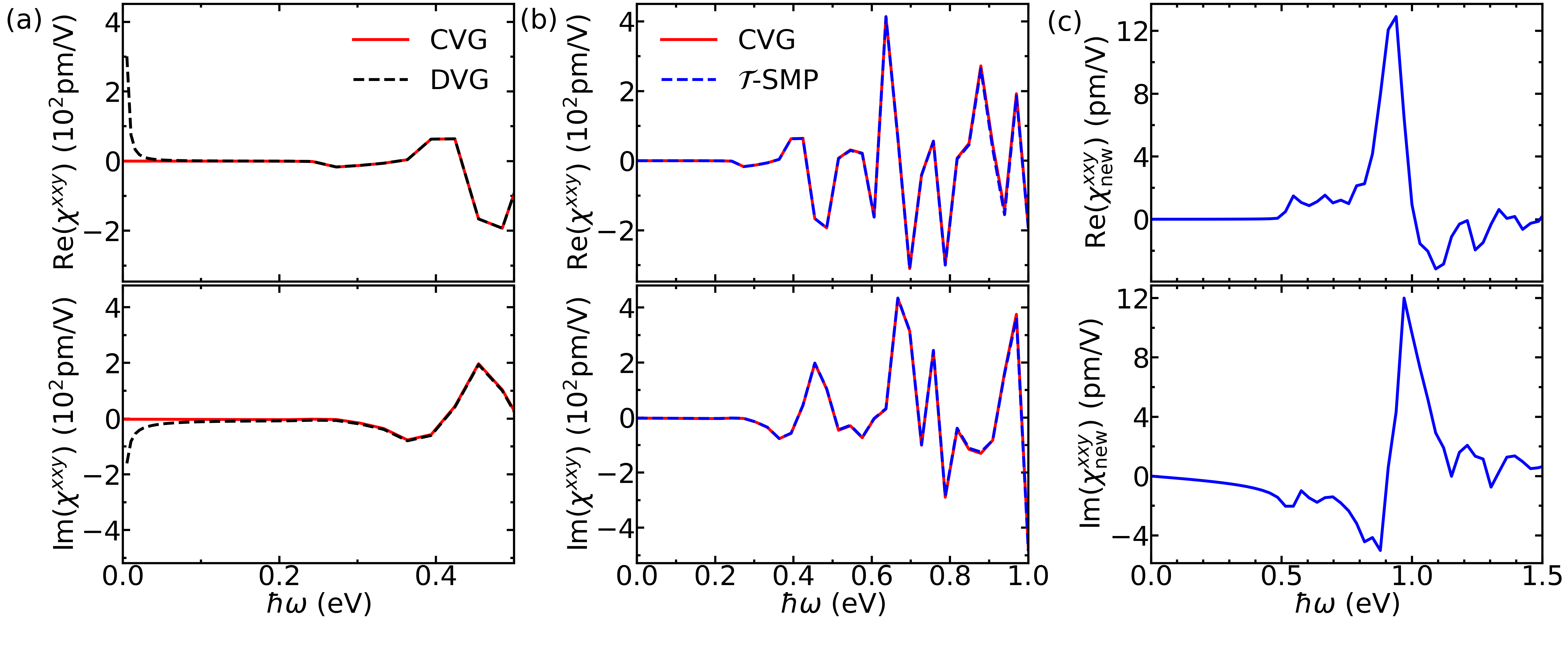}
    \caption{SHG of bilayer AB$^{\prime}$ stacking CrI$_3$ along $xxy$ direction. Upper and lower pannels of (a) are the real and imaginary part of SHG, respectively. The black dashed line shows the results of the previous divergent formulae marked as 'DVG', the red solid line shows the results of complete and modified formulae marked as 'CVG'. (b) shows the comparison between the complete and modified formulae ('CVG', red solid line) and the conventional used but not complete formulae ($\mathcal{T}$-SMP, blue dashed line). (c) shows the difference between them, where the blue line is the value of the new term.} 
    \label{Fig5}
\end{figure*}

Figure \ref{Fig5}(a) shows the real and imaginary parts of SHG calculated using the DVG and the CVG expressions.
As the CVG expression is derived from the DVG expression, results from both expressions are exactly the same away from the low-frequency region ($>$ 0.1 eV). 
However, in the low-frequency region ($0-0.1$ eV), both the real and imaginary parts calculated from the DVG expression exhibit divergent problem which is absent in the CVG expression. 
Additionally, in the static limit, the CVG expression converges to exactly zero which is consistent with our symmetry analysis in Sec. \ref{sec:symmetry}.

Figure \ref{Fig5}(b) shows the real and imaginary parts of SHG calculated using the $\mathcal{T}$-SMP and the CVG expressions. 
As the SHG response in bilayer CrI$_3$ is gigantic, the difference between the two expressions is hardly observed in Fig. \ref{Fig5}(b).  
Detailed analyses in Fig. \ref{Fig5}(c) demonstrates that the differences between the two expressions, which is $\chi_{\rm new}$, are roughly 10 pm/V. 
The materials and frequency range in which $\chi_{\rm new}$ is considerable is subject to further investigation with more magnetic materials.

\section{Conclusion}\label{sec:conclusion}
In summary, we have given a thorough derivation of second-order magneto-optical effects, including rectification current and second-order susceptibility using density matrix perturbation method.
Especially, the formalism of rectification current conductivity is compatible with both non-degenerate and degenerate bands and is applicable to both non-magnetic and magnetic materials. 
Additionally, the second-order susceptibility with $\omega^{-1}$ and $\omega^{-2}$ terms, which suffers from low-frequency divergent problem, numerical instability and is dropped conventionally, has been reformulated into a full-frequency convergent form without any symmetry assumptions. 
Furthermore, we have implemented the above formulae to first-principles calculations, which has not been done for real materials as far as we know, using both orthogonal and non-orthogonal basis methods. 
Two $\mathcal{PT}$-symmetric magnetic materials, bilayer AFM MnBi$_2$Te$_4$ and CrI$_3$ are used as examples. 
We confirmed that our formulae for rectification current including the MIC and MSC are truly invariant under U(2)-gauge freedom.
Meanwhile, for SHG in magnetic materials, we found noticeable modifications of SHG responses from the divergent term. 
With the vast variety of magnetic materials, it is expected that our reformulated formalism can greatly facilitate and advance the theoretical understandings on non-linear magneto-optical effects.

\begin{acknowledgments}
We thank Chong Wang and Shiqiao Du for helpful discussion. 
This work was supported by the Basic Science Center Project of NSFC (Grant No. 51788104), 
the Ministry of Science and Technology of China (Grants 2018YFA0307100, and 2018YFA0305603), the National Science Fund for Distinguished Young Scholars (Grant No. 12025405), the National Natural Science Foundation of China  (Grant No. 11874035), and the Beijing Advanced Innovation Center for Future Chip (ICFC). M. Y. is supported by Shuimu
Tsinghua Scholar Program and Postdoctoral International Exchange Program.
\end{acknowledgments}

\appendix

\newpage

\section{Basic Hamiltonian, current operator and polarization operator}\label{appendix Basic}
\subsection{Basic Hamiltonian}\label{appendix Basic Hamiltonian}
The Hamiltonian of a semiconductor in an electromagnetic field, whose magnetic field is negligible and the electric field is treated in the long-wavelength limit, can be written as \cite{Sipe_2000}
\begin{equation} \label{Hamiltonian}
\begin{aligned}
    \hat{H}(t)&=\int \tilde{\psi}^{\dagger}(\mathbf{x})\left[H_{0}-e \mathbf{x} \cdot \mathbf{E}(t)\right] \widetilde{\psi}(\mathbf{x}) d \mathbf{x},
\end{aligned}
\end{equation}
where $H_{0}$ is the unperturbated single particle Hamiltonian in the coordinate representation with Bloch eigenfunctions $\psi_{n_{\mu}}(\mathbf{k} ; \mathbf{x})$ and eigenvalues $\hbar \omega_{n}(\mathbf{k})$, 
and $\tilde{\psi}(\mathbf{x})=\sum_{n}\sum_{\mu} \int d \mathbf{k}  \psi_{n_{\mu}}(\mathbf{k} ; \mathbf{x})\hat{a}_{n_{\mu}}(\mathbf{k})$ is the field operator with the annihilation operator for Bloch state as $\hat{a}_{n_{\mu}}(\mathbf{k})$.
$n$ and $\mu$ are the band indices and $e=-|e|$.
The interaction of electric field and electrons is represented using the length gauge by the $-e \mathbf{x} \cdot \mathbf{E}(t)$ term while the results are equivalent to the velocity gauge formula  \cite{Ventura_gauge_equivalence,Sipe_1995}. 
The long-wavelength limit is valid as long as the wavelength of light is much larger than the length scale of interests.

Combining Eq. \eqref{eq:x_matrix} and Eq. \eqref{Hamiltonian}, we get
\begin{equation}
    \begin{aligned}
    \hat{H}(t)=&\int d\mathbf{k}\hbar\omega_{n_{\mu}}(\mathbf{k})\hat{a}^{\dagger}_{n_{\mu}}(\mathbf{k})\hat{a}_{n_{\mu}}(\mathbf{k})\\
    &-e\int d\mathbf{k}\mathbf{E}(t)\cdot\mathbf{r}_{n_{\mu} m_{\nu}}(\mathbf{k})\hat{a}^{\dagger}_{n_{\mu}}(\mathbf{k})\hat{a}_{m_{\nu}}(\mathbf{k})\\
    &-e\int d\mathbf{k}\mathbf{E}(t)\cdot\hat{a}^{\dagger}_{n_{\mu}}(\mathbf{k})i\partial_{\mathbf{k}}\hat{a}_{n_{\mu}}(\mathbf{k})\\
    &-e\int d\mathbf{k}\mathbf{E}(t)\cdot\mathbf{A}_{n_{\mu}}(\mathbf{k})\hat{a}^{\dagger}_{n_{\mu}}(\mathbf{k})\hat{a}_{n_{\mu}}(\mathbf{k})\,.
    \end{aligned}
\end{equation}

The intraband electric current according to Eq. \eqref{eq:jtot} is 
\begin{equation}    \label{jintra}
\begin{aligned}
    {\hat{j}}_{\rm intra}^{a}=&{e} \int d\mathbf{k} v_{n_{\mu} n_{\mu}}^{a}(\mathbf{k})\hat{a}_{n_{\mu}}^{\dagger}(\mathbf{k}) \hat{a}_{n_{\mu}}(\mathbf{k})\\
    &-\frac{e^2}{\hbar}  \int d\mathbf{k} E^{b}r^{b}_{n_{\mu}m_{\nu};a}\hat{a}_{n_{\mu}}^{\dagger}(\mathbf{k}) \hat{a}_{m_{\nu}}(\mathbf{k})\\
    &-\frac{e^2}{ \hbar}  \int d\mathbf{k} E^{b}\Omega_{n_{\mu}}^{ab}\hat{a}_{n_{\mu}}^{\dagger}(\mathbf{k}) \hat{a}_{n_{\mu}}(\mathbf{k})
\end{aligned}
\end{equation}
and the interband polarization of Eq.(\ref{eq:jtot}) is
\begin{equation} \label{Pinter}
    \hat{P}^{a}_{\rm inter}=e \int d \mathbf{k}  r^a_{n_{\mu} m_{\nu}}(\mathbf{k}) \hat{a}_{n_{\mu}}^{\dagger}(\mathbf{k}) \hat{a}_{m_{\nu}}(\mathbf{k}) \,,
\end{equation}
where $r^{b}_{n_{\mu}m_{\nu};a}=\partial_a r^{b}_{n_{\mu}m_{\nu}}-ir^{b}_{n_{\mu}m_{\nu}}(A^{a}_{n_{\mu}}-A^{a}_{m_{\nu}})$ is the U(1)-covariant derivative \cite{Sipe_2000}.

\subsection{Dynamics of density operators}\label{appendix: Basic density matrix}
The matrix element of density operator is  
\begin{equation}
    \langle n_{\lambda} \mathbf{k}|\hat{\rho}| m_{\mu} \mathbf{k}^{\prime} \rangle=\langle \hat{a}^{\dagger}_{m_{\mu}}(\mathbf{k}^{\prime})\hat{a}_{n_{\lambda}}(\mathbf{k}) \rangle \,,
\end{equation}
and the initial density operator without perturbation is
\begin{equation} \label{zeroth density}
    \hat{\rho}_0=\frac{1}{Z}e^{-\beta {H}_{0}}
\end{equation}
with the matrix element $\langle n_{\lambda} \mathbf{k}|\hat{\rho}_0|m_{\mu} \mathbf{k}^{\prime} \rangle=f_{n_{\lambda}}\delta_{nm}\delta_{\lambda \mu}\delta(\mathbf{k}-\mathbf{k}^{\prime}),$ where the prefactor $f_{n_{\lambda}}$ is the band occupation number.

Switching on illumination at $t=-\infty$, the Heisenberg equation of motion of the density operator matrix element reads \cite{Boyd_book,Baltz_first_shift_current1981,Kraut_first_shift_current1979}
\begin{equation}\label{eq: equation of motion}
    \frac{d}{d t}\hat{\rho}_{n_{\lambda} m_{\mu}}(t)=\frac{1}{i\hbar}[H(t),\hat{\rho}(t)]_{n_{\lambda} m_{\mu}}- \frac{\hat{\rho}_{n_{\lambda} m_{\mu}}(t)-\hat{\rho}_{n_{\lambda} m_{\mu}}(0)}{\tau_{n_{\lambda} m_{\mu}}},
\end{equation}
where the last term is a phenomenological term describing scattering processes of electrons with the relaxation time ${\tau_{n_{\lambda}m_{\nu}}}$.
Most frequently, the relaxation time is assumed to be independent of the initial and final states, and therefore the subscript of $\tau$ is dropped.

Expanding the density operator in power of the electric field as $\hat{\rho}(t)=\hat{\rho}_0+\hat{\rho}_1(t)+\hat{\rho}_2(t) + \cdots $ and inserting it into Eq. (\ref{eq: equation of motion}), the equation of motion of density operator can be rewritten as  
\begin{equation}
    i\hbar\frac{d \hat{\rho}_{i+1}(t)}{d t}=[H_0,\hat{\rho}_{i+1}(t)]+[H^{\prime},\hat{\rho}_{i}(t)]-i\hbar \frac{\hat{\rho}_{i+1}(t)}{\tau},
\end{equation}
which can be solved using the iteration method.
The first-order term of the density operator matrix element is
\begin{equation} \label{first order density operator}
    (\hat{\rho}_{1})_{n_{\lambda} \mathbf{k}, m_{\mu} \mathbf{k}^{\prime}}=\frac{eE^{b}_{\beta}e^{-i\omega_{\beta}t}}{\hbar}\frac{f_{m n}r^b_{n_{\lambda}m_{\mu}}(\mathbf{k})}{\omega_{n m}-\omega_{\beta}-i/\tau}\delta(\mathbf{k}-\mathbf{k}^{\prime})\,,
\end{equation}
and the second-order term is
\begin{equation} \label{second order density operator}
    (\hat{\rho}_2)_{n_{\lambda} \mathbf{k}, m_{\mu} \mathbf{k}^{\prime}}=\frac{-e^2}{\hbar^2}\frac{(\Gamma_2)_{n_\lambda \mathbf{k}, m_\mu \mathbf{k}^{\prime}}\delta(\mathbf{k}-\mathbf{k}^{\prime})}{\omega_{n m}(\mathbf{k})-\omega_{\Sigma}-i/\tau}E_{\beta}^{b}E_{\gamma}^{c}e^{-i\omega_{\Sigma}t}
\end{equation}
with
\begin{equation}
\begin{aligned}
    &(\Gamma_2^L)_{n_{\lambda} \mathbf{k}, m_{\mu} \mathbf{k}}\\
    &=\frac{f_{l n}(\mathbf{k})r_{n_{\lambda} l_{\nu}}^{b}(\mathbf{k})r_{l_{\nu} m_{\mu}}^{c}(\mathbf{k})}{\omega_{n l}(\mathbf{k})-\omega_{\beta}-i/\tau}-\frac{f_{m l}(\mathbf{k})r_{l_{\nu} m_{\mu}}^{b}(\mathbf{k})r_{n_{\lambda} l_{\nu}}^{c}(\mathbf{k})}{\omega_{l m}(\mathbf{k})-\omega_{\beta}-i/\tau}\\
&\qquad \qquad -i\left[\frac{f_{m n}(\mathbf{k})r_{n_{\lambda} m_{\mu}}^{b}(\mathbf{k})}{\omega_{n m}(\mathbf{k})-\omega_{\beta}-i/\tau}\right]_{;c} \,.
\end{aligned}
\end{equation}
In derivation of (\ref{second order density operator}), we have assumed that the system is a gapped system, in other words, $\partial_ {\mathbf{k}}f_{n}(\mathbf{k})=0$.

\subsection{First- and second-order response functions}\label{appendix: Basic responses}
The zeroth-order polarization response can be obtained through Eq. \eqref{eq:x_matrix} and Eq. \eqref{zeroth density} as
\begin{equation}
    P^a_0=-e\int[d\mathbf{k}]f_{n}A^{a}_{n_{\mu}}(\mathbf{k})\,,
\end{equation}
which reproduces the result of modern theory of polarization.

The first-order electric susceptibility is acquired by combining Eq. \eqref{eq:x_matrix} and Eq. \eqref{first order density operator} as 
\begin{equation} \label{first order susceptibility}
\begin{aligned} 
    \chi_{1}^{ab}(-\omega;\omega)&=\frac{e^2}{\hbar}\int [d \mathbf k]\frac{f_{nm}r^a_{n_{\mu}m_{\nu}}r^b_{m_{\nu}n_{\mu}}}{\omega_{mn}-\omega-i/\tau}\\
    &-\frac{e^2}{i\hbar \omega} \int [d \mathbf k] f_{n}[\frac{\partial}{\partial k_a} A_{n_{\mu}}^b -\frac{\partial}{\partial k_b} A_{n_{\mu}}^a ]\,.
\end{aligned}
\end{equation}
While the first term provides a finite susceptibility, the second term shows a $\omega^{-1}$ divergence in the static limit ($\omega\to0$) which indicates the presence of a DC current in the static limit.
Using the continuity relationship, the first-order electric current conductivity is 
\begin{equation} 
\begin{aligned} 
    {\sigma}^{ab}_{1}(-\omega;\omega)=&\frac{-ie^2\omega}{\hbar}\int [d \mathbf k]\frac{f_{nm}r^a_{n_{\mu}m_{\nu}}r^b_{m_{\nu}n_{\mu}}}{\omega_{mn}-\omega-i/\tau}\\
    &+\frac{e^2}{\hbar}\int [d \mathbf k] f_{n}\Omega_{n_{\mu}}^{ab}\,.
\end{aligned}
\end{equation}
In the static limit, the first term vanishes and the second term is just the anomalous Hall current \cite{Nagaosa_review_AHC}. 
For two-dimensional insulators with non-zero Chern number, this term contributes a quantized anomalous Hall current.

For second order responses, $\langle j^{a}_{\rm intra}(t) \rangle^{(2)}$ has been given in the main text and the full expression of $\langle{P}_{\rm inter}^{a} (t)\rangle^{(2)}$ is given by 

\begin{equation} \label{eq: full Pinter}
\begin{aligned}
    &\langle\mathbf{P}_{\rm inter}(t)\rangle^{(2)}\\
    =& \frac{e^3}{2\hbar^2}\int [d \mathbf{k}] \frac{[\ldots]E_{\beta}^{b} E_{\gamma}^{c} e^{-i \omega_{\Sigma} t}}{\omega_{ln}-\omega_{\Sigma}-i/\tau}\\
    =& \frac{e^{3}}{2\hbar^{2}} \int [d \mathbf{k}]\frac{r_{n_{\mu} l_{\nu}}^{a}}{\omega_{l n}-\omega_{\Sigma}-i/\tau}\left[\frac{f_{n m} r_{l_{\nu} m_{\lambda}}^{c} r_{m_{\lambda} n_{\mu}}^{b}}{\omega_{m n}-\omega_{\beta}-i/\tau} \right.\\
    &\left. -\frac{f_{m l} r_{l_{\nu} m_{\lambda}}^{b} r_{m_{\lambda} n_{\mu}}^{c}}{\omega_{l m}-\omega_{\beta}-i/\tau} + \frac{i f_{nl} r_{l_{\nu} n_{\mu} ; c}^{b}}{\omega_{l n}-\omega_{\beta}-i/\tau} \right. \\
    &\left. - \frac{i f_{nl} r_{l_{\nu} n_{\mu}}^{b} \Delta_{l n}^{c}}{\left(\omega_{l n}-\omega_{\beta}-i/\tau\right)^{2}} + (b\beta \leftrightarrow c\gamma) \right]E_{\beta}^{b} E_{\gamma}^{c} e^{-i \omega_{\Sigma} t}.\\
\end{aligned}
\end{equation}

\section{Alternative derivation of the U(2)-invariant formulation}\label{appendix: another way of U(2)}
Starting from Eq. \eqref{eq:x_matrix} and considering the position operator matrix between degenerate bands in the intraband term, an alternate division of the interband and intraband position operator matrix elements gives  
\begin{equation}  
\begin{aligned}
    \langle n_{\mu} \mathbf{k}|\mathbf{\hat{x}}_{\rm intra}|m_{\nu} \mathbf{k^{\prime}}\rangle=&\delta_{nm}\delta_{\mu\nu}[\mathbf{A}_{n_{\mu} }(\mathbf{k})+i\frac{\partial}{\partial\mathbf{k}}\delta(\mathbf{k}-\mathbf{k^{\prime}})]\\
    &+\delta_{nm}(1-\delta_{\mu\nu})\mathbf{r}_{n_{\mu} m_{\nu}}(\mathbf{k}) \,,
\end{aligned} \label{eq:x_intra_new}
\end{equation}
and
\begin{equation}  
\begin{aligned}
    \langle n_{\mu} \mathbf{k}|\mathbf{\hat{x}}_{\rm inter}|m_{\nu} \mathbf{k^{\prime}}\rangle=&(1-\delta_{nm})\mathbf{r}_{n_{\mu}m_{\nu}}(\mathbf{k}) \,.
\end{aligned}
\end{equation}
The newly added second term in Eq. \eqref{eq:x_intra_new} is the origin of the diverging $P_{\rm inter}(\omega_\Sigma)$ in the $\omega_\Sigma \to 0$ limit. 
Following the standard treatment of the main text, we retrieved the same results regardless of how the position operator matrix elements are divided.
In the new division, DC current is only from $J_{\rm intra}$ term regardless of the presence of the degeneracy condition.

\section{Full expression of SHG and the static limit}\label{appendix: SFG and SHG}
In the main text, we focused on the divergent terms in the second-order susceptibility, while the full expressions consist of both the non-divergent term originated from Eq. \eqref{eq: full Pinter} and the `apparent' diverging terms in Eq. \eqref{eq:nodvg}.
In addition, as SHG is frequently measured and computed, we provided the full expression of SHG as
\begin{equation}
    \chi_{2}^{abc}(-2\omega;\omega,\omega)=\chi_{e}^{abc}(-2\omega;\omega,\omega)  + \chi_i^{abc}(-2\omega;\omega,\omega) \label{eq: full SHG}
\end{equation}
with  
\begin{equation}
    \begin{split}
        &\chi_e^{abc}(-2\omega;\omega,\omega)\\
        =&\frac{e^3}{2\hbar^2} \int [d\mathbf{k}] \sum^{\prime} \frac{r_{n_{\mu} m_{\nu}}^a ( r_{m_{\nu} l_{\lambda}}^{b} r_{l_{\lambda} n_{\mu}}^{c} + r_{m_{\nu} l_{\lambda}}^{c} r_{l_{\lambda} n_{\mu}}^{b} )}{\omega_{ln}-\omega_{ml}} \\
        & \times\left(\frac{2f_{nm}}{\omega_{mn}-2\omega} + \frac{f_{ln}}{\omega_{ln}-\omega} + \frac{f_{ml}}{\omega_{ml}-\omega}\right)\\
    \end{split} \label{eq:chi_e}
\end{equation}
is purely the contribution from interband process and  
\begin{equation}
    \begin{split}
        &\chi_i^{abc}(-2\omega;\omega,\omega)\\
        =& \frac{i e^{3}}{2 \hbar^{2}} \int [d\mathbf{k}]  f_{nm} \left[ \frac{2(I_{nm}^{cab} + I_{nm}^{bac})}{\omega_{mn}(\omega_{mn}-2 \omega)} + \frac{I_{mn}^{cba} + I_{mn}^{bca}}{\omega_{mn}(\omega_{mn}-\omega)} \right. \\
        & + \frac{ (g_{nm}^{ab}-\frac{i}{2}\Omega_{nm}^{ab}) \Delta_{mn}^c + (g_{nm}^{ac}-\frac{i}{2}\Omega_{nm}^{ac}) \Delta_{mn}^b }{\omega_{mn}^2} \\
        &\times \left(\frac{1}{\omega_{mn}-\omega}-\frac{4}{\omega_{mn}-2\omega}\right)\\ &\left. - \frac{I_{mn}^{acb} + I_{mn}^{abc} }{2\omega_{mn}(\omega_{mn}-\omega)}-\frac{\Delta^{a}_{mn} g_{nm}^{bc} }{2\omega^{2}_{mn}(\omega_{mn}-\omega)} \right]
    \end{split} \label{eq:chi_i}
\end{equation}
is the contribution of the mixed interband and intraband processes where the last two terms are $\chi_{\rm dvg}$.
Again, the small imaginary part $i/\tau$ in the denominator is not written out explicitly. The $\sum^{\prime}$ in Eq. \eqref{eq:chi_e} means that the in the summation, $n,m,l$ are not equal to each other. 
In addition to the new term derived in Eq. \eqref{eq:SHG_nodvg}, our full expression also satisfies that each term is gauge-covariant regardless of the degeneracy condition. 
This is achieved by restricting $n, m, l$ not equal to each other in the three-band summation term and rearranging the rest terms not obeying this restriction (terms including connections in degenerate sub space) with the U(1)-covariant derivative to construct the U(2)-covariant derivative.

The SHG susceptibility described in Eq. \eqref{eq:chi_e} and Eq. \eqref{eq:chi_i} is made up of terms of the form $\int \frac{A}{\omega_{mn}-\omega -i/\tau}$. 
The $\mathcal{T}$-even part of the susceptibility $\chi_{\rm N}$ takes the form of $\int \frac{\Re{A}}{\omega_{mn}-\omega -i/\tau} $ and the $\mathcal{T}$-odd part $\chi_{\rm M}$ takes the form of $\int \frac{\Im{A}}{\omega_{mn}-\omega -i/\tau}$. 
Therefore, in the limit $\tau \to \infty $, the principal part and $\delta$-function part of the non-magnetic $\chi_{\rm N}$ are purely real and imaginary respectively, while the real and imaginary parts are opposite in the magnetic $\chi_{\rm M}$ \cite{Pershan_1963,Shen_1989_MSHG}.

The SHG susceptibility at the static limit is an important criterion for the application of non-linear crystals and we rearranged the static susceptibility in a form that directly represent the Kleinman's symmetry as
\begin{equation}
    \begin{split}
        &\chi_{e}^{abc}(0;0,0) \\
        =& \frac{e^3}{6\hbar^2} \int [dk] \sum^{\prime} P(abc) \Re{r_{n_\mu m_\nu}^a r_{m_\nu l_\lambda}^b r_{l_\lambda n_\mu}^c} \\
        & \times \frac{\omega_m f_{nl} + \omega_p f_{mn} + \omega_n f_{lm}}{\omega_{mn} \omega_{ln} \omega_{ml}}
    \end{split}
\end{equation}
and
\begin{equation}
    \chi_{i}^{abc}(0;0,0) = -\frac{e^3}{4\hbar^2} \int [dk]  \frac{f_{nm}}{\omega_{mn}^2} P(abc) \Im{I_{nm}^{abc}},
\end{equation}
where $P(abc)$ denotes full permutation of indices $a,b,c$.

\section{The relationship between energy dissipation and susceptibility} \label{appendix:dissipation}
In $\mathcal{T}$-symmetric systems, the imaginary part of susceptibility which only contains the $\delta$-function part characterizes the dissipation of the electromagnetic field energy, while the real part does not contain $\delta$-function terms.
In $\mathcal{T}$-breaking systems, the susceptibility is complex in general, and the imaginary part of the susceptibility is not related to dissipation. 
For example, the power of dissipation in the first order gives 
\begin{equation}
\begin{aligned}
    &\mathbf{E}\cdot \frac{d\mathbf{P}_1}{dt}\\ 
    =&\sum_{\omega} i\omega\left[-\chi^{ab}_{1}(-\omega;\omega)+\chi^{ba}_{1}(\omega;-\omega)\right] E^{a}_{-\omega}E^{b}_\omega + c.c. \\
    =&\sum_{\omega}\frac{\pi e^2\omega}{\hbar}\int [d\mathbf{k}] f_{nm} \left[ r_{nm}^{a}r_{mn}^{b}E^{a}_{-\omega} E^b_\omega + c.c.\right] \\
    &\times \delta(\omega_{mn}-\omega)\,.
\end{aligned}
\end{equation}
Therefore, in $\mathcal{T}$-breaking systems, the dissipation is not related to the imaginary part of $\chi_1$, rather, it depends only on the $\delta$-function part which is not purely imaginary. 

\section{Band structure of bilayer CrI$_3$} \label{appendix:band}
\begin{figure}\includegraphics[width=\linewidth]{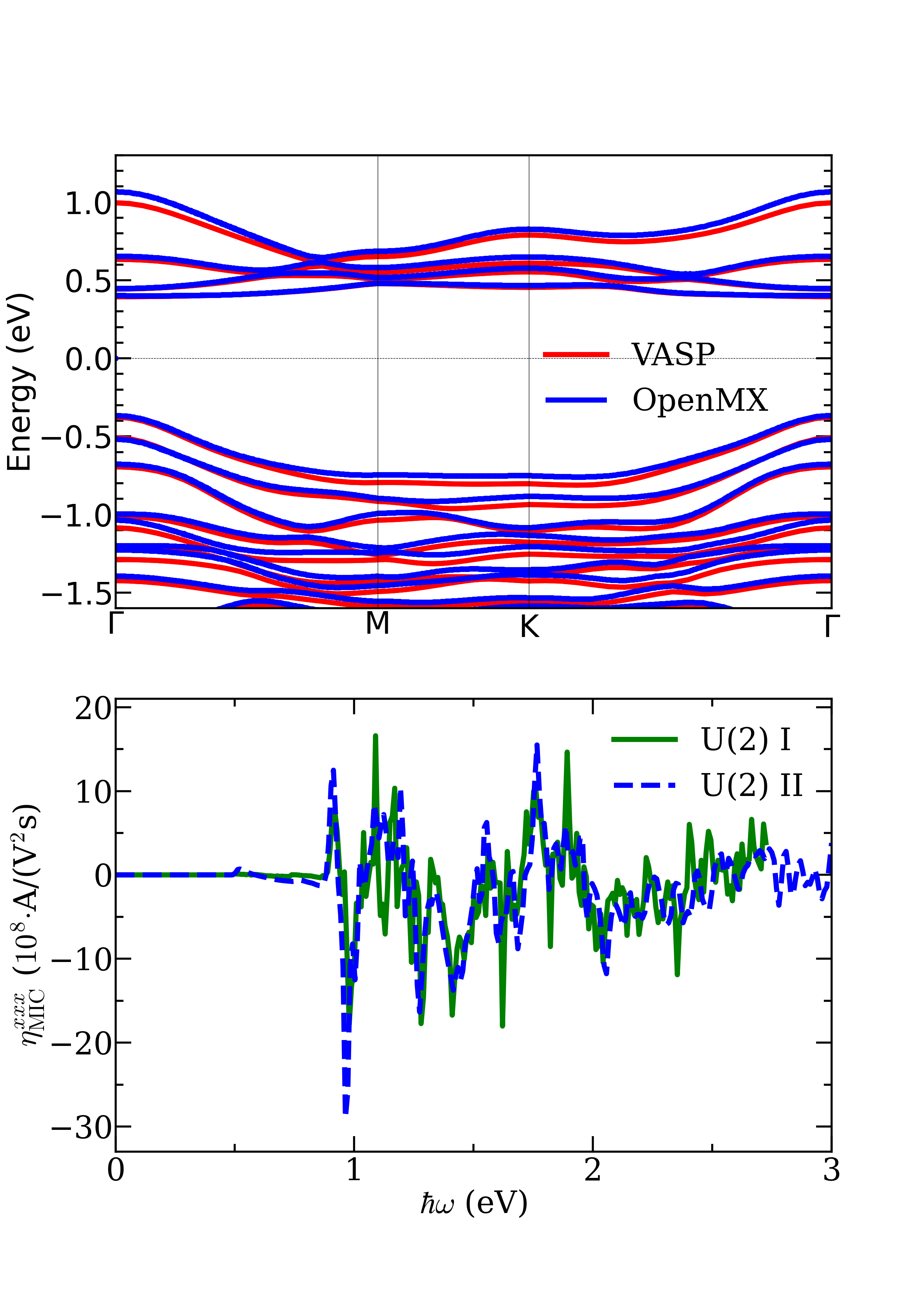}
 \caption{Band structure and MSC conductivity of bilayer CrI$_3$ calculated from that from route \uppercase\expandafter{\romannumeral2} compared with route \uppercase\expandafter{\romannumeral1}, both using U(2) invariant formulae with first-principles packages VASP and OpenMX. We have applied a scissor operation to the results of OpenMX to increase the gap to the same value as that of VASP.} 
 \label{Fig7}
\end{figure}

Fig. \ref{Fig7} shows the band structure and MSC conductivity calculated from route \uppercase\expandafter{\romannumeral1} and route \uppercase\expandafter{\romannumeral2} using the U(2)-invariant formula, where the electronic structures are obtained from different first-principles packages while other parameters are essentially the same. 
The electronic structures calculated from the two packages are slightly different in many aspects including bandgap values and band dispersion as shown in Fig. \ref{Fig7}(a).
For instance, the band gap is 0.89\,eV from VASP and 0.51\,eV from OpenMX.
Even though the scissor operator has been applied to align the bandgap in Fig. \ref{Fig4}(b), different features in MSC are still observed re-emphasizing that the optical current is sensitive to the electronic structure and it is difficult to compare results from different packages.

\nocite{*}
%

\end{document}